\documentclass[12pt,a4paper]{article}
\usepackage{amssymb,amsmath,amsfonts}
\usepackage{hyperref}
\usepackage{color, xcolor}
\usepackage[dvipsnames]{xcolor}
\usepackage[T1]{fontenc}
\usepackage{cite}
\hypersetup{
    colorlinks=true,
    linkcolor=blue,
    filecolor=magenta,      
    urlcolor=cyan,
    pdftitle={Overleaf Example},
    pdfpagemode=FullScreen,
    }
    
\definecolor{blue-violet}{rgb}{0.54, 0.17, 0.89}
\definecolor{PineGreen}{cmyk}{0.92, 0, 0.59, 0.25}
\definecolor{OliveGreen}{cmyk}{0.64, 0, 0.95, 0.40}
\definecolor{RawSienna}{cmyk}{0, 0.72, 1, 0.45}
\definecolor{Gray}{cmyk}{0, 0, 0, 0.50}
\definecolor{MidnightBlue}{cmyk}{0.98, 0.13, 0, 0.43}
\definecolor{Orange}{cmyk}{0, 0.61, 0.87, 0}
\definecolor{LimeGreen}{cmyk}{0.50, 0, 1, 0}
\definecolor{Green}{cmyk}{1, 0, 1, 0}

\usepackage[makeroom]{cancel}
\usepackage[normalem]{ulem}

\numberwithin{equation}{section}

\begin{document}

\title{\bf Holographic Weyl Anomaly and Kounterterms in AdS gravity}

\author{Giorgos Anastasiou$^{1}$\thanks{georgios.anastasiou@uai.cl}\;, 
Jahaira Bonifacio-Chavez$^{2,3}$\thanks{{jahaira.bonifacio93@gmail.com}}\;,
Olivera Miskovic$^{4,5}$\thanks{olivera.miskovic@pucv.cl}\;,\\
and Rodrigo Olea$^{4}$\thanks{rodrigo\_olea\_a@yahoo.co.uk}
\bigskip\\
{\small\it $^1$Departamento de Ciencias, Facultad de Artes Liberales, Universidad Adolfo Ibañez,}\vspace{-0.1cm}\\
{\small\it Av.~Diagonal Las Torres 2640, Peñalolen, Chile.}\vspace{-0.1cm}\\
{\small\it $^2$Instituto de Ciencias Exactas y Naturales, Universidad Arturo Prat,}\vspace{-0.1cm}\\
{\small\it Playa Brava 3256, 1111346, Iquique, Chile.}\vspace{-0.1cm}\\
{\small\it $^3$Facultad de Ciencias, Universidad Arturo Prat,}\vspace{-0.1cm}\\
{\small\it Avenida Arturo Prat Chacón 2120, 1110939, Iquique, Chile.}\vspace{-0.1cm}\\
{\small\it $^4$Instituto de F\'\i sica, Pontificia Universidad Cat\'olica de Valpara\'\i so,}\vspace{-0.1cm}\\ 
{\small\it Avenida Universidad 330, Curauma, Valpara\'{\i}so, Chile.}\vspace{-0.1cm}\\
{\small\it $^5$DISAT, Politecnico di Torino, Corso Duca degli Abruzzi, 24, 10129 Torino, Italy.}}

\date{}

\maketitle

\begin{abstract}
The addition of Kounterterms to Einstein gravity leads to a finite action for asymptotically anti-de Sitter (AdS) spaces with a conformally flat boundary. In that sense, it provides a partial renormalization for AdS gravity when compared to standard holographic techniques, where the mismatch is given in terms of nontrivial conformal properties of the boundary. On the other hand, this method has the clear advantage that the variation of the action has a closed form in an arbitrary dimension.

In this work, it is shown how to extract holographic information on conformal anomalies from the variation in $(2n+1)$-dimensional Einstein-AdS plus Kounterterms. Remarkably enough, a considerable part of the Weyl anomaly can be worked out for any odd dimension.
\end{abstract}

\newpage

\section{Introduction}

One of the key features which realizes the duality between anti-de Sitter (AdS) gravity in a spacetime and a Conformal Field Theory at its boundary,  is the existence of a  conformal structure in the asymptotic region~\cite{Maldacena:1997re, Gubser:2002tv, Witten:1998qj}. Indeed, the dual boundary field theory  lives on a curved background metric $g_{(0)ij}$, which is a representative of a conformal class of metrics. In the light of the analysis of residual asymptotic symmetries, radial diffeomorphisms induce Weyl transformations for the metric at the conformal boundary~\cite{Imbimbo:1999bj}, fact that was long anticipated in Ref.~\cite{Brown:1986nw}. 

The dynamics of AdS gravity in $d+1$ dimensions is governed by the Einstein-Hilbert action 
\begin{equation}\label{QCGaction}
I_{\mathrm{EH}}=-\frac{1}{16\pi G }\int\limits_{\mathcal{M}}d^{d+1}x\sqrt{-g} 
\,(R-2\Lambda) \,,  
\end{equation}
where the cosmological constant $\Lambda$ is negative and given by $\Lambda=-d(d-1)/2 \ell^2$ in terms of the AdS radius $\ell$. This bulk functional, by taking arbitrary variations on the metric, produces the field equations
\begin{equation}\label{eom}
 \mathcal{E}_{\mu\nu} :=  R_{\mu\nu} - \frac{1}{2}g_{\mu\nu}R + \Lambda g_{\mu\nu}  = 0\,.
\end{equation}
Generic solutions to this gravity theory are Einstein spaces which are of constant, negative curvature in the asymptotic region. It was shown in Ref.~\cite{AST_1985__S131__95_0} that the metric for any asymptotically AdS (AAdS) space may be put in the form
\begin{equation}
ds^2 = \frac{\ell^2}{z^2} \left(dz^2 + \bar{g}_{ij} \left(z,x\right) dx^i dx^j\right)\,,
\label{metric}
\end{equation}
what reflects a divergence of order two at the conformal boundary. The boundary metric $\bar{g}_{ij}(z,x)$  
 admits a regular power-series expansion. In principle, from a kinematic standpoint, there may be odd powers in $z$ in the Fefferman-Graham (FG) expansion~\eqref{FGexpansion}, however, these modes are vanishing when Einstein equations are imposed. In doing so, the expansion is given by
\begin{equation}
\bar{g}_{ij}=
g_{(0)ij}+\frac{z^{2}}{\ell^2}\,g_{(2)ij}+\frac{z^{4}}{\ell^4}\,g_{(4)ij}+\cdots +\frac{z^{d}}{\ell^d}\left(\,g_{(d)ij}+h_{(d)ij} \log \frac{z^2}{\ell^2} \right)+\cdots \,.
\label{FGexpansion}
\end{equation}
The log term only appears in even-dimensional boundaries. 
For an arbitrary metric at the conformal boundary $g_{(0)ij}$, FG frame is a realization of the asymptotic condition on the curvature
\begin{equation} \label{RAdSFG}
R^{\alpha \beta}_{\mu \nu}=-\frac{1}{\ell^2}\delta^{\alpha \beta}_{\mu \nu}+\mathcal{O}(z^2)\,.
\end{equation}
Furthermore, the higher-order coefficients can be determined, up to the normalizable order $\mathcal{O}\left(z^d\right)$, as a function of the holographic data $g_{\left(0\right)ij}$, by the asymptotic solution of the equations of motion. Indeed, for the next-to-leading order coefficient, one gets
\begin{equation} \label{g2}
g_{(2)ij} = -\ell ^{2}\mathcal{S}_{(0)ij} \,,
\end{equation}
in terms of the Schouten tensor~\cite{deHaro:2000vlm,Skenderis:2002wp} associated to the holographic metric $g_{0)ij}$, that is,
\begin{equation}
\mathcal{S}_{(0)ij}=\frac{1}{d-2}\left( \mathcal{R}_{(0)ij}-\frac{\mathcal{R}_{(0)}}{2(d-1)}\,g_{0)ij}\right) \,.  \label{Schouten}
\end{equation}%
At the $z^4$ order in the FG expansion, the identification of the coefficient $g_{4)ij}$ in terms of four-derivative covariant objects of the holographic data~\cite{Anastasiou:2020zwc} is given by
\begin{equation} \label{g4}
g_{(4)ij} = \frac{\ell^4}{4(d-4)}\,\left[(d-4)\mathcal{S}^i_{(0)k}\mathcal{S}^k_{(0)j}-\mathcal{B}^i_{(0)j}\right]\,.
\end{equation}
Here, $\mathcal{B}^i_{(0)j}$ is the Bach tensor which, in turn, is defined as
\begin{equation}   \mathcal{B}^i_{(0)j}=\nabla^{k}_{(0)}\mathcal{C}^i_{(0)jk} + \mathcal{S}_{(0)}^{kl}\mathcal{W}^i_{(0)kjl}\,,
\end{equation}
in terms of the divergence of the Cotton tensor
\begin{equation}
\mathcal{C}^{i}_{(0)jk} =\nabla_{(0)k}\mathcal{S}^{i}_{(0)j}-\nabla_{(0)j}\mathcal{S}^{i}_{(0)k}\,.
\label{Cotton}
\end{equation}
The appearance of Schouten, Weyl, Cotton and Bach tensors may be thought as a reflection of underlying conformal structures at the boundary, what is natural in the context of AdS/CFT correspondence. This idea is reinforced by the expression for the FG mode at $z^6$ order in the expansion~\cite{Jia:2021hgy}
\begin{equation}
g_{\left(6\right)ij} =\frac{\ell^6}{24\left(d-4\right) \left(d-6\right)} \left[4\left(d-6\right) \mathcal{B}_{\left(0\right)ik} \mathcal{S}^k_{\left(0\right)j}-\mathcal{O}^{\left(6\right)}_{ij}\right] \,,
\end{equation}
where $\mathcal{O}^{\left(6\right)}_{ij}$ is the six-derivative obstruction tensor
\begin{align}
\mathcal{O}^{\left(6\right)}_{ij}&= \nabla_{k}^{\left(0\right)}\nabla^{\left(0\right)k} \mathcal{B}_{\left(0\right)ij}+2\mathcal{B}^{kl}_{\left(0\right)}\mathcal{W}_{\left(0\right)ikjl} - 4 \mathcal{S}_{\left(0\right)} \mathcal{B}_{\left(0\right)ij} + 2 \left(d-4\right) \left(2 \mathcal{S}^{kl}_{\left(0\right)} \nabla^{\left(0\right)}_l \mathcal{C}_{\left(0\right) (ij)k} \right. \nonumber \\
&  \left. + \mathcal{C}_{\left(0\right) (ij)}^{\qquad k} \nabla^{\left(0\right)}_{k} \mathcal{S}_{\left(0\right)} 
 -\mathcal{C}^{s \quad k}_{\left(0\right) i}\mathcal{C}_{\left(0\right) kjs} + \nabla^{\left(0\right)k} \mathcal{S}^{m}_{\left(0\right)(i} \mathcal{C}_{\left(0\right) j)mk}+ \mathcal{S}^{k}_{\left(0\right)m} \mathcal{S}^{ml}_{\left(0\right)}\mathcal{W}_{\left(0\right)ikjl}\right)\,.
\end{align}
The Weyl tensor, $\mathcal{W}
_{(0)kl}^{ij}$, is given in terms of curvatures associated to the metric at the conformal boundary as
\begin{equation}
\mathcal{W}
_{(0)kl}^{ij}=\mathcal{R}_{(0)kl}^{ij}-\mathcal{S}_{(0)k}^{i}\delta
_{l}^{j}+\mathcal{S}_{(0)l}^{i}\delta _{k}^{j}+\mathcal{S}_{(0)k}^{j}\delta _{l}^{i}-\mathcal{S}_{(0)l}^{j}\delta
_{k}^{i}\,.  \label{W}
\end{equation}
The asymptotic form of the FG metric allows one to have control on the infrared divergences in the gravitational action and its variation. Indeed, the systematic construction of counterterms that remove these divergent terms relies on the FG expansion. However, in order to restore covariance at the boundary, it is required to write down the counterterms $L_{\rm ct}$ as covariant expressions of the full boundary metric $h_{ij}$ in Gauss-normal coordinates \cite{deHaro:2000vlm}
\begin{equation}
     ds^2=N^2(z)dz^2+h_{ij}(z,x)\,dx^idx^j\,.
     \label{metricGN}
 \end{equation}
 The fact the series of counterterms is intrinsic is dictated by a variational problem based on Dirichlet boundary conditions for the metric $h_{ij}$, what is achieved by the addition of the Gibbons-Hawking term, that is,
\begin{equation}
I_{\rm{ren}}
=  I_{\rm EH} -\dfrac{1}{8\pi G }\int\limits_{\partial \mathcal{M}}d^{d}x\,\sqrt{-h
}\,K+\int\limits_{\partial \mathcal{M}}d^{d}x\,L_{ct}(h,\mathcal{R},\mathcal{\nabla R})\,.
\end{equation}%
In the above relation, $K$ is the trace of the extrinsic curvature, which is defined in the current coordinate frame as 
  \begin{equation} \label{Kij}
  K_{ij}=-\frac{1}{2N} \,\partial_zh_{ij}\,.
 \end{equation}
AdS gravity in low enough dimension requires just a few  counterterms to be regular. In higher dimensions, however, the series turns far  more complicated~\cite{Balasubramanian:1999re, Emparan:1999pm}
\begin{eqnarray}
8\pi G  L_{ct}&=&\frac{d-1}{\ell }\sqrt{-h}+\frac{\ell \sqrt{-h}}{%
2(d-2)}\mathcal{R}+\frac{\ell ^{3}\sqrt{-h}}{2(d-2)^{2}(d-4)}\left( \mathcal{%
R}^{ik}\mathcal{R}_{ik}-\frac{d}{4(d-1)}\mathcal{R}^{2}\right) \notag \\
&+&\frac{\ell ^{5}\sqrt{-h}}{(d-2)^{3}(d-4)(d-6)}\left( \frac{3d-2}{4(d-1)}%
\mathcal{RR}^{ik}\mathcal{R}_{ik}-\frac{d(d+2)}{16(d-1)^{2}}\mathcal{R}%
^{3} \right. \notag \\
&-& \left. 2\mathcal{R}^{ik}\mathcal{R}^{jl}\mathcal{R}_{ijkl}-
\frac{d}{4(d-1)}\nabla _{l}\mathcal{R}\nabla ^{l}\mathcal{R}+\nabla ^{l}\mathcal{R}^{ik}\nabla _{l}\mathcal{R}_{ik}\right) +\cdots,
\end{eqnarray}
where $\mathcal{R}_{kl}^{ij}=\mathcal{R}_{kl}^{ij}(h)$ is the Riemann curvature of the boundary metric. The Ricci tensor $\mathcal{R}_{j}^{i}$ and the Ricci scalar $\mathcal{R}$ are then obtained by taking traces of the Riemann tensor. From a practical point of view, the above formula is useful to perform computations in particular gravitational solutions in a different coordinate frame, e.g., Schwarzschild-like coordinates.

Counterterms can be equivalently expressed in a different basis made by tensors used in Conformal Calculus, as~\cite{Anastasiou:2020zwc}
\begin{align}
8\pi G  L_{ct} &= \frac{d-1}{\ell }\sqrt{-h}+\frac{\ell \left(d-1\right)\sqrt{-h}}{%
\left(d-2\right)}\mathcal{S} - \frac{\ell^3 \sqrt{-h}}{2\left(d-4\right)} \delta^{ik}_{jl}\mathcal{S}^{j}_{i}\mathcal{S}^{l}_{k} \notag \\
&+ \frac{\ell^{5}\sqrt{-h}}{\left(d-2\right)\left(d-4\right)\left(d-6\right)} \left(\mathcal{S}^i_j \mathcal{B}^j_i +\frac{d-4}{2} \delta^{ikm}_{jln}\mathcal{S}^{j}_{i}\mathcal{S}^{l}_{k} \mathcal{S}^{n}_{m} \right) +\mathcal{O}\left(\mathcal{R}^4\right)\,.
\end{align}
The Schouten and Bach tensors in the present case, are defined in terms of the induced metric $h_{ij}$.

By definition, the quasilocal stress tensor is obtained as the variation of the total action with respect to $h_{ij}$, that is,
\begin{equation}
 \delta I_{\rm ren} = \frac{1}{2}  \int\limits_{\partial \mathcal{M}}d^dx\sqrt{-h}\,\, T^{ij}\delta h_{ij}\,.
\end{equation}%
This energy-momentum tensor is the sum of the canonical momentum (associated to a radial foliation)
\begin{equation}
 \pi^{ij}[h] = \frac{1}{8\pi G }\left(K^{ij}- K h^{ij}\right)\,,
\end{equation} 
plus a contribution due to boundary counterterms
\begin{equation}
 T^{ij}[h] = \pi^{ij}+\frac{2}{\sqrt{-h}}\frac{\delta L_{ct}}{\delta h_{ij}}\,.
\end{equation} 
The component $(iz)$ of the Einstein equation guarantees  the conservation of the boundary stress tensor and, therefore, the existence of quasilocal charges
\begin{equation}
    {Q}[\xi] = \int\limits_{\Sigma_\infty}d^{d-1}y\sqrt{|\gamma|}\,T^i_j\,\xi^j\, u_i\,,
\end{equation}
for a given set of isometries $\{\xi^i\}$ at the boundary. This is an integral defined on a codimension-2 surface $\Sigma_\infty$ endowed with local coordinates $\{y^a\}$ and a metric $\gamma_{ab}$. The vector $u_i$ is the timelike normal to $\Sigma_\infty$.

The above formula requires no reference spacetime as a background in order to measure geometric fluctuations from. As a matter of fact, background-independent charges play a key role in the gauge/gravity duality. In order to illustrate this point, one may review the computation of the energy for 
Schwarzschild-AdS black hole in five dimensions, where the mass appears with an extra, constant term
\begin{equation}
    {Q}[\partial_t] = M+E_0\,,\hspace{1cm}E_0=\frac{\ell^2}{32\pi G }\,.
\end{equation}
Within the framework of AdS/CFT correspondence, an interesting duality for vacuum/Casimir energy was proposed in Ref.~\cite{Balasubramanian:1999re}. In fact, the Casimir energy for a four-dimensional CFT  on $\mathbb{R}\times \mathbb{S}^3$ takes the generic form
\begin{equation}
    E_{\rm{Casimir}}=\frac{1}{960\ell}\left( 4n_s+17 n_f+88n_v \right)
\end{equation}
in terms of the number of scalar, fermion, and vector fields. The proper use of the AdS/CFT dictionary relates the constants of the bulk gravity to the ones of the boundary field theory, that is, $\ell^3/G=2N^2/\pi$. The vacuum energy, then, gets a suggestive form
\begin{equation}
    E_0=\frac{3N^2}{16\ell}\,,
\end{equation}
what allows for the identification of the dual boundary CFT.
The matching is met in the large $N$ limit for 
\begin{equation} \label{fieldSYM}
n_{s}=6(N^{2}-1)\,,\qquad n_{f}=2(N^{2}-1)\,,\qquad n_{v}=N^{2}-1\,,
\end{equation}
that is, $\mathcal{N}=4$ $SU(N)$ super Yang-Mills theory.

Vacuum energy in higher odd dimensions $D=2n+1$ is~\cite{Emparan:1999pm}
\begin{equation} \label{E0}
E_{0}=(-1)^{n}\frac{(2n-1)!!^{2}}{(2n)!}\frac{\mathrm{Vol}(\mathbb{S}^{2n-1})}{8\pi G }\ell^{2n-2}\,.
\end{equation}
In a $ d$-dimensional CFT, classical Weyl transformations give rise to the trace of the stress tensor. At the quantum level, the theory is governed by the renormalized effective action $W[g_{(0)}]$. This is the finite generating functional obtained from the quantum effective action after removing UV divergences which, in general, is not invariant under Weyl transformations in even dimensions. Since the variation of $W$ is related to the CFT stress tensor $T^{ij}$ through the relation
\begin{equation}
\delta W[g_{(0)}]=\dfrac{1}{2}\int d^{d}x\sqrt{-g_{(0)}}\,\langle
T^{ij}(x)\rangle \,{\delta g_{(0)ij}\,,}  \label{generating functional}
\end{equation}
an infinitesimal Weyl transformation ${\delta }_{\sigma }{g_{(0)ij}=2\sigma
\,g_{(0)ij}}$ leads to
\begin{equation}
\delta _{\sigma }W[g_{(0)}]=\int d^{d}x\,\sqrt{-g_{(0)}}\ \sigma \mathcal{A} \,,
\end{equation}
so that the Weyl or trace anomaly is given by
\begin{equation}
\mathcal{A}=\langle T^{ij}\rangle \,g_{(0)ij}=\langle T^i_{\ i}\rangle \,.  
\end{equation}
A local Weyl anomaly $\mathcal{A}(x)$  exists only in even dimensions, $d=2n$, where it is a function of curvature invariants. In odd dimensions, any local Weyl anomaly vanishes, although non-local (or
boundary) anomalies may still exist~\cite{Deser:1976yx}.

Early work in four-dimensional conformal anomalies can be found in Refs.~\cite{Capper:1974ic,Capper:1975ig,Deser:1976yx,Duff:1977ay,Birrell:1982ix,Duff:1993wm,Deser:1996na}. Deser and Schwimmer in Ref.~\cite{Deser:1993yx} conjectured a classification of conformal anomalies in even dimensions. This classification was later proven by Boulanger in Refs.~\cite{Boulanger:2007ab, Boulanger:2007st}. Indeed, the general form of the conformal anomaly consists of three types of contributions
\begin{equation}
\mathcal{A}=\mathcal{A}_{\mathrm{A}}+\mathcal{A}_{\mathrm{B}}
+ \mathcal{A}_{\mathrm{C}}\,. \label{A}
\end{equation}
The first one is the type A anomaly, 
\begin{equation}
\mathcal{A}_{\mathrm{A}}= a \,\mathcal{E}_{2n}\,, \label{typeA}
\end{equation} 
which is always proportional to the Euler density,
\begin{eqnarray}
    \mathcal{E}_{2n}&=&\frac{1}{2^{n}}\,\delta _{i_{1}\cdots i_{2n}}^{j_{1}\cdots
j_{2n}}\mathcal{R}_{(0)j_{1}j_{2}}^{i_{1}i_{2}}\cdots \mathcal{R}_{(0)j_{2n-1}j_{2n}}^{i_{2n-1}i_{2n}} \notag \\
&=& \mathrm{Pf}(\mathcal{R}_{(0)})\,,
\label{Euler}
\end{eqnarray}
where $\mathrm{Pf}(\cdots)$ stands for the Pfaffian, i.e., the totally antisymmetric product of curvature tensors.

On the other hand, the type B anomaly comes from the logarithmic UV divergences. Namely, the presence of the log scale introduces a dimensionful parameter, which breaks Weyl invariance. This anomaly is then given by local Weyl invariant densities $I_{(k)}$ 
\begin{equation}
\mathcal{A}_{\mathrm{B}}= -\sum\limits_k \tilde{c}_k \,I_{(k)}\,,\
\end{equation}
which can be expressed in terms of the Weyl tensor and its derivatives~\cite{zbMATH05710186, Boulanger:2018rxo}. The explicit form of these conformal invariants is known only up to eight dimensions~\cite{Boulanger:2004zf}. 

In turn, type C anomaly is given in terms of a total derivative,
\begin{equation}
\mathcal{A}_{\mathrm{C}}=\nabla_i V^i\,,
\end{equation}
which is a cohomologically trivial contribution to $\mathcal{A}$. Its presence  does not pose a physical obstruction to gauge invariance, as it may be eliminated by a local counterterm added to the effective action.\footnote{
A number of articles have further elaborated on the classification by Deser and Schwimmer. In particular, they have reviewed the argument on total derivatives in the anomaly being  cohomologically trivial.
Indeed,  anomalies of the form $\mathcal{A}=\sigma dA'$ are trivial only if the primitive $A'$ itself
is BRST/Weyl invariant \cite{Boulanger:2007st}. The distinction stems from the fact the covariant derivative cannot generate Weyl-invariant
quantities, without including a compensating field. Therefore, the Weyl-covariant derivative is a key ingredient
in the improved classification of anomalies by Boulanger.

This observation justifies the inclusion of two boundary terms in the 12-term basis of conformal invariants in eight dimensions \cite{Boulanger:2004zf,Boulanger:2025oli}, as they are cohomologically nontrivial. They should carry independent information on central charges in the conformal anomaly.}

In order to employ the AdS$_{d+1}$/CFT$_d$ correspondence, on the gravity side, the holographic stress tensor has to be obtained by a suitable rescaling of the quasilocal energy-momentum tensor 
\begin{equation}
    \langle T^i_j \rangle = \lim_{z \to 0} \left( \frac{\ell^{d}}{z^{d}} \, T^i_j [h] \right) \,.
\end{equation}
The corresponding generating functional of the boundary CFT is identified with the on-shell gravitational action in AdS, after removing IR divergences,
\begin{equation}
    W[g_{(0)}]=I_{\mathrm{ren}}[g_{(0)}] \,,
\end{equation}
such that the resulting finite renormalized bulk action plays the role of the renormalized effective action of the CFT. This identification allows one to compute the holographic Weyl anomaly
\begin{equation}
\delta _{\sigma }I^{\rm ren}_{2n+1}=\int\limits_{\partial \mathcal{M}} d^{2n}x\,\sqrt{-g_{(0)}}\,\sigma \,\langle T_{\
i}^{i}\rangle=\int\limits_{\partial \mathcal{M}} d^{2n}x\,\sqrt{-g_{(0)}}\,\sigma \,\mathcal{A}\,. \label{defA}
\end{equation}
In the case of five-dimensional AdS gravity, the computation of holographic 1-point function gives rise to the Weyl anomaly, which takes the explicit form~\cite{Henningson:1998gx}
\begin{equation}
\mathcal{A}=\dfrac{\ell ^{3}}{128\pi G }\,\left( \mathcal{E}_{4}-\mathcal{%
W}^{2}\right) \,,  \label{4D anomaly}
\end{equation}%
where $\mathcal{E}_{4}$ is the Euler density~\eqref{Euler} (type A anomaly), while $\mathcal{W}^{2}\equiv \,\mathcal{W}%
^{ijkl}\mathcal{W}_{ijkl}$ is the only conformal invariant in four dimensions term, given by a  quadratic term in the boundary Weyl tensor $\mathcal{W}%
_{kl}^{ij}(g_{(0)})$.

As AdS/CFT prescription naturally arises in the context of type IIB superstring theory~\cite{Maldacena:1997re} compactified on AdS$_{5}\times \mathbb{S}^{5}$, one is able to relate
10-dimensional parameters to the ones in five dimensions as $G ^{(10)}=G \,\mathrm{Vol}(\mathbb{S}^{5})=\ell ^{5}\pi ^{3}G $. On the other hand, Newton's constant in ten dimensions is expressed as $G ^{(10)}=8\pi ^{6}g_{\mathrm{s}}^{2}\,\alpha ^{\prime 4}$.  Furthermore, the near-horizon
geometry of $N$ coincident D3-branes gives the AdS radius as $\ell ^{4}=4\pi g_{\mathrm{s}}N\,\alpha'$. Setting $\alpha'=1$, substitution yields the familiar relation $G =\frac{\pi \ell ^{3}}{2N^{2}}$, and therefore
the anomaly~\eqref{4D anomaly} in AdS$_5$ adopts the form
\begin{equation}
\mathcal{A}=\dfrac{N^{2}}{64\,\pi ^{2}}\,\left( \mathcal{E}_{4}-\mathcal{W}%
^{2}\right) \,.  \label{5D result}
\end{equation}
On the CFT$_{4}$ side, the trace anomaly of a general four--dimensional
conformal field theory with $n_{s}$ real scalars, $n_{f}$ Dirac fermions and
$n_{v}$ vector fields is given by~\cite{Duff:1993wm}
\begin{equation}
\mathcal{A}_{\mathrm{CFT}}=\frac{1}{5760\,\pi ^{2}}\,(n_{s}+11n_{f}+62n_{v})
\,\mathcal{E}_{4}-\frac{1}{1920\,\pi ^{2}}\,(n_{s}+6n_{f}+12n_{v}) \, \mathcal{W}^{2}\,.
\end{equation}
In particular, for $\mathcal{N}=4$ super Yang-Mills $SU(N)$ gauge theory, with the field content given by Eq.\eqref{fieldSYM},
one has
\begin{equation}
\mathcal{A}_{\mathrm{CFT}}=\frac{N^{2}-1}{64\,\pi ^{2}}\,\left( \mathcal{E}_{4}-\,\mathcal{W}^{2}\right) \,,
\end{equation}%
and, therefore, showing agreement with Eq.~\eqref{5D result} in the large $N$ limit.

In a similar fashion, for the case AdS$_7$/CFT$_6$, the anomaly reads~\cite{Jia:2021hgy}
\begin{equation}
\mathcal{A} = \langle T^i_i \rangle = - \frac{\ell^5}{64 \pi G_N} \left( \delta^{i_1 i_2 i_3}_{j_1 j_2 j_3} \mathcal{S}^{j_1}_{\left(0\right)i_1} \mathcal{S}^{j_2}_{\left(0\right)i_2} \mathcal{S}^{j_3}_{\left(0\right)i_3} +\mathcal{S}^{j}_{\left(0\right)i}\mathcal{B}^{i}_{\left(0\right)j}\right) \,.
\end{equation}
This expression can be integrated by parts such that, up to a total derivative term, the 6D conformal anomaly can be cast in the form\footnote{The type B anomaly in this case is a given combination of the three conformal invariants in the basis  by Osborn and Stergiou. One of the conformal invariants, though, which is not polynomial in the Weyl tensor, requires a total derivative in order to be a local Weyl invariant (i.e., including boundary terms)~\cite{Osborn:2015rna}.}
\begin{equation}
\mathcal{A} =  - \frac{\ell^5}{3072 \pi G_N} \left[ \mathcal{E}_{6} - \mathrm{Pf}(\mathcal{W}_{(0)}) - \frac{3}{2} \delta^{i_1 \ldots i_5}_{j_1 \ldots j_5} \mathcal{W}^{j_1 j_2}_{ \left(0\right) i_1 i_2} \mathcal{W}^{j_3 j_4}_{ \left(0\right) i_3 i_4} \mathcal{S}^{j_5}_{\left(0\right)i_5} - 24\, \mathcal{C}_{\left(0\right)j k}^{i} \mathcal{C}_{\left(0\right)i}^{jk}\right] \,.
\label{7Danomalykarydas}
\end{equation}
On general grounds, a holographic derivation of the  central charge $a$ in an arbitrary gravity theory is expected to be forbiddingly involved in high enough dimensions.
However, in Ref.~\cite{Imbimbo:1999bj}, a universal prescription on how to obtain the coefficient of the type A anomaly was provided, based on a near-boundary analysis of AAdS spaces. The general formula, in $d=2n$ boundary dimensions, is given by \footnote{The result for the  central charge $a$ in Ref.~\cite{Imbimbo:1999bj} appears here multiplied by $\ell$, as required by dimensional analysis.}
\begin{equation}
a=\frac{\ell ^{2n+1}}{2^{2n}n!^{2}}\,\mathcal{L}|_{_{\mathrm{AdS}}}\,,
\end{equation}
where $\mathcal{L}|_{_{\mathrm{AdS}}}$ is the Lagrangian density  evaluated on global AdS spacetime. 
In particular, for Einstein gravity, the Ricci scalar takes the value $R|_{_{\mathrm{AdS}}}=-\frac{2n(2n+1)}{\ell^2}$, such that 
\begin{equation}
\mathcal{L}_{\mathrm{EH}}|_{_{\mathrm{AdS}}}=\frac{1}{16\pi G }\,\frac{4n}{\ell ^{2}}\,.  \label{b0}
\end{equation}
Then, the central charge associated to the type A anomaly is expected to be
\begin{equation}
a=\frac{1}{16\pi G }\frac{n\ell ^{2n-1}}{2^{2n-2}n!^{2}}\,,
 \label{a}
\end{equation}
for any CFT dual to Einstein gravity. From all possible conformal invariants, in this case, the evidence up to nine dimensions indicates that the type-B anomaly adopts the form
\begin{equation}
\mathcal{A}_{\mathrm{B}}=-c\, \mathrm{Pf}(\mathcal{W}_{(0)})+\cdots\,.
\end{equation}
that is, the totally anti-symmetric product of maximal powers of the Weyl tensor. In Einstein gravity, holographic computations lead to $c=a$. The rest of the terms are polynomial in the Weyl and Schouten tensors, and higher-derivative contributions.

\section{Kounterterms in anti-de Sitter gravity}

Standard holographic techniques rely on the asymptotic solution of the field equations, order by order in $z$ until the holographic mode, which can be regarded as a perturbative method around a constant-curvature background in Eq.\eqref{RAdSFG}. The limitation of such a perturbative approach is given by finding the explicit form of higher-order coefficients in the FG expansion. As suggested by Eqs.~\eqref{g2} and~\eqref{g4}, these terms should also be expressible in conformal tensors (Weyl tensor and Weyl-covariant derivatives of it) such that they vanish for conformally flat bulk manifolds~\cite{Skenderis:1999nb}. Thus, the technical difficulty comes from the possibility of writing down the expansion of the metric as tensors of interest in Conformal Calculus beyond a certain order in the radial coordinate $z$.

The addition of extrinsic counterterms as surface terms provides an alternative to Holographic Renormalization that circumvents this obstacle. This proposal, dubbed Kounterterms,  was introduced in Refs.~\cite{Olea:2005gb,Olea:2006vd}, and considers a boundary term
\begin{equation}
I_{\rm KT}=I_{\mathrm{EH}}-c_{d}\int\limits_{\partial \mathcal{M}}d^{d}x\,B_{d}(h,K,\mathcal{R})\,,  \label{Ireg}
\end{equation}
whose dependence on extrinsic curvature manifestly spoils a variational principle based on a Dirichlet boundary condition for the boundary metric $h_{ij}$. However, this method is  consistent with a Dirichlet problem in terms of the holographic metric $g_{(0)ij}$ and, therefore, compatible with a holographic picture of AdS gravity.

In $D=2n$ dimensions, the corresponding surface term is given in terms of a single parametric integral
\begin{align}
B_{2n-1}& =-2n\sqrt{-h}\int\limits_{0}^{1}dt\ \delta _{ i_{1}\cdots
i_{2n-1} }^{ j_{1}\cdots j_{2n-1} }K_{j_{1}}^{i_{1}}%
\left( \frac{1}{2}\mathcal{R}%
_{j_{2}j_{3}}^{i_{2}i_{3}}-t^{2}K_{j_{2}}^{i_{2}}K_{j_{3}}^{i_{3}}\right)
\times \cdots \notag \\
& \hspace{2cm}\cdots \times \left( \frac{1}{2}\mathcal{R}%
_{j_{2n-2}j_{2n-1}}^{i_{2n-2}i_{2n-1}}-t^{2}K_{j_{2n-2}}^{i_{2n-2}}K_{j_{2n-1}}^{i_{2n-1}}\right),
\end{align}
which can be thought of as a polynomial of the extrinsic and intrinsic curvatures whose coefficients can be worked out from the binomial expansion. The overall factor
\begin{equation} \label{c2n-1}
c_{2n-1} =\frac{1}{16\pi G}\frac{\left(-1\right)^n\ell ^{2n-2}}{n(2n-2)!}\,,
\end{equation}
is singled out by the fact that the total action is zero for global AdS, which is a minimal consistency check in order to reproduce black hole thermodynamics in even-dimensional AdS gravity.

As for Einstein-AdS gravity in odd dimensions $D=2n+1$, the Kounterterms series is given by the double parametric integral
\begin{eqnarray}
B_{2n} &=&-2n\sqrt{-h}\int\limits_{0}^{1}dt\int\limits_{0}^{t}ds\,\delta
_{i_{1}\cdots i_{2n}}^{j_{1}\cdots j_{2n}}\,K_{j_{1}}^{i_{1}}\delta
_{j_{2}}^{i_{2}}\left( \frac{1}{2}\,\mathcal{R}%
_{j_{3}j_{4}}^{i_{3}i_{4}}-t^{2}K_{j_{3}}^{i_{3}}K_{j_{4}}^{i_{4}}+\frac{%
s^{2}}{\ell ^{2}}\,\delta _{j_{3}}^{i_{3}}\delta _{j_{4}}^{i_{4}}\right)
\times \cdots   \notag \\
&&\cdots \times \left( \frac{1}{2}\,\mathcal{R}%
_{j_{2n-1}j_{2n}}^{i_{2n-1}i_{2n}}-t^{2}K_{j_{2n-1}}^{i_{2n-1}}K_{j_{2n}}^{i_{2n}}+%
\frac{s^{2}}{\ell ^{2}}\,\delta _{j_{2n-1}}^{i_{2n-1}}\delta
_{j_{2n}}^{i_{2n}}\right) \,,  \label{B2n}
\end{eqnarray}
with a corresponding coupling constant that takes the form
\begin{eqnarray}
c_{2n} &=&-\frac{1}{16\pi G }\frac{(-\ell ^{2})^{n-1}}{n(2n-1) !}\,\left[ \int\limits_{0}^{1}dt\,\left( 1-t^{2}\right) ^{n-1}\right] ^{-1}  \notag \\
&=&\frac{1}{16\pi G }\frac{\left(-1\right)^n \ell ^{2n-2}}{2^{2n-2}n\left(
n-1\right) !^{2}}\,.  \label{c2n}
\end{eqnarray}
Kounterterms may be employed to achieve a finite action for the Euclidean form of Schwarzschild-AdS black hole
\begin{equation}
    ds^2=f^2(r)d\tau^2+f^{-2}(r)dr^2+r^2d\Omega^2_{d-1}\,.
\end{equation}
Here, the Euclidean time is $\tau=i t$, $d\Omega^2_{d-1}$ is the line element of the sphere $\mathbb{S}^{d-1}$, and the metric function is given by
\begin{equation}
    f^2(r)=1-\frac{2\omega_dG M}{r^{d-2}}+\frac{r^2}{\ell^2}\,,
\end{equation}
in terms of the (Hamiltonian) mass $M$ and a dimension-dependent numerical factor $\omega_d$ necessary to recover the black hole entropy as a quarter of the area of the horizon.

Direct evaluation of the bulk action depicts the typical divergent behavior of the volume for AAdS spaces
\begin{equation}
TI_{\rm EH}^{\rm (E)}=\frac{(d-2)}{(d-1)}M-TS+\lim_{r\rightarrow \infty }\frac{\mathrm{Vol}(\mathbb{S}^{d-1})}{8\pi G }\frac{r^{d}}{\ell ^{2}}\,,
\end{equation}%
and an incorrect factor for the mass to play the role of the internal energy of the thermodynamic system. Standard definitions for thermodynamic variables as Hawking temperature $T$ and entropy $S$ have been taken.

In turn, the Euclidean version of the Kounterterms produces the missing factor in the mass plus a divergent term which cancels the infinities in the above result, i.e.,
\begin{equation}
T\,c_{d}\int\limits_{\partial \mathcal{M}}d^dx\,B_{d}^{\rm (E)}=\frac{M}{(d-1)}+E_{0}-\lim_{r\rightarrow \infty }\frac{\mathrm{Vol}(\mathbb{S}^{d-1})}{8\pi G }\frac{r^{d}}{\ell ^{2}}\,.
\end{equation}
Therefore, Kounterterm method is able to reproduce the standard thermodynamic relations for Einstein-AdS black holes, in terms of the Helmholtz free energy $F$
\begin{equation} \label{QSR}
F=U-TS
\end{equation}%
as long as the internal energy is identified as the mass plus the vacuum energy $E_0$~\cite{Emparan:1999pm, Olea:2006vd}
\begin{equation}
U=M+E_{0}\,.
\end{equation}
The derivation of a Quantum Statistical Relation~\eqref{QSR} without resorting to any background-substraction procedure, in any bulk dimension, is an interesting consequence of the addition of extrinsic counterterms.

On more general grounds, one can explore to what extent the proposal of Kounterterms could reproduce the  results of renormalization given by standard techniques in the context of AdS/CFT correspondence. As a matter of fact, in Ref.~\cite{Anastasiou:2020zwc} an explicit comparison between these two schemes was performed for a generic holographic metric $g_{(0)ij}$, leading to 
\begin{equation} \label{mismatch}
I_{\rm{KT}}=I_{\text{HR}}-\frac{\ell ^{3}}{64\pi G (d-2)(d-4)}%
\int\limits_{\partial \mathcal{M}}\sqrt{-h}\,\mathcal{W}^2(h) +\cdots\,.
\end{equation}
The mismatch is expressed by a series of terms with an increasing number of derivatives which would eventually capture different conformal properties of the boundary. It starts with a boundary Weyl-squared contribution, followed by a six-derivative term, which may depend on the boundary Schouten, Weyl, Cotton and Bach tensors. The latter term would be divergent --and should be duly renormalized by additional counterterms-- in bulk dimensions equal to or greater than six.

The difference between the extrinsic and intrinsic counterterms is zero for most AAdS spaces (Schwarzschild-AdS, Kerr-AdS, black strings), which explains the fact that Kounterterms method is able to recover the thermodynamics associated to these gravitational solutions. Indeed, for any asymptotically conformally flat (ACF) spacetime in AdS gravity, the results obtained by Holographic Renormalization can be recovered. For boundaries with nontrivial conformal properties, as in the case of topological black holes or gravitational instantons, the gravitational action will be divergent.

The mismatch in Eq.~\eqref{mismatch} can then be seen as an ambiguity of the Kounterterms prescription, as the corresponding couplings~\eqref{c2n-1} and~\eqref{c2n} are fixed by demanding finiteness for global AdS spacetimes. That means that there might be additional boundary terms that depend on the bulk Weyl tensor whose couplings cannot be fixed by the previous criterion.

In order to match the standard counterterms at the boundary of even-dimensional AdS gravity, a different scheme, known as Conformal Renormalization, has been recently discussed. It considers the consistent embedding Einstein theory in Conformal Gravity, defined as a linear combination of conformal invariants in the bulk. This linear combination is such that the corresponding field equations (of higher order in derivatives) accept Einstein spaces as a generic solution. This mechanism of renormalization was originally inspired by the four-dimensional case, where Einstein spaces are always a well-defined sector of Conformal Gravity (whose action is given by the square of the Weyl tensor)~\cite{Anastasiou:2016jix}. Holographic boundary conditions imply the cancelation of higher-derivative modes in Conformal Gravity such that the asymptotic form of the metric is that of Einstein gravity~\cite{maldacena2011einsteingravityconformalgravity}. At the level of the action, the Einstein-AdS action appears augmented by the Gauss-Bonnet term, which is able to generate the correct counterterms from its equivalent form as a surface term~\cite{Miskovic:2009bm}.

In six spacetime dimensions, the construction relies on the Lu-Pang-Pope Conformal Gravity, which was shown to contain Schwarzschild-AdS black hole as a solution~\cite{Lu:2013hx}. It was later proven in Ref.~\cite{Anastasiou:2020mik} that the theory possesses the entire sector of Einstein spaces. This fact is of particular relevance, as conformal invariants may include surface terms which ensure the Weyl invariance also at the boundary. These boundary contributions, dictated by conformal invariance of the total action, when taken to the Einstein sector of the theory, eliminate the extra terms in Eq.~\eqref{mismatch}. A similar procedure, by only at the level of the gravity action and not for the field equations, has been recently proposed for Einstein-Gauss-Bonnet gravity with negative cosmological constant in six dimensions~\cite{Anastasiou:2025usa}.

In eight bulk dimensions, a Conformal Gravity theory with an Einstein sector has been recently reported in Ref.~\cite{Boulanger:2025oli}. Verifying whether conformal invariance encodes the renormalization of Einstein-AdS action also in that case is a matter of current investigation.

In contrast, in odd-dimensional Einstein-AdS gravity, the partial renormalization given the addition of Kounterterms cannot be amended in a similar fashion as in the even-dimensional case. In fact, in odd bulk dimensions (but just for $D=4k-1$), the only conformal invariants available are gravitational Chern-Simons terms, which are parity-odd structures~\cite{Boulanger:2018rxo}. In that respect, they cannot contain an Einstein sector, whose action is parity preserving. In any case, the existence of conformal anomaly would spoil a possible link to a Weyl-invariant gravitational action.

On the other hand, the fact that this extrinsic renormalization is able to reproduce the value of the vacuum/Casimir energy~\eqref{E0}, relevant to establish gauge/gravity relations, and that $E_0$ can be conveniently rewritten as
\begin{equation}
    E_0=(2n-1)!\,c_{2n}\mathrm{Vol}(\mathbb{S}^{2n-1})\,,
\end{equation}
in terms of the coupling $c_{2n}$, strongly suggests a connection to holographic quantities. In particular, it opens the question on --to what extent--  holographic anomalies may be read off within this scheme.

\section{Variation of the total action: pre-holographic arguments}
\label{Sec:Variation}

In order to work out the different contributions to the holographic conformal anomaly, the on-shell variation of the action~\eqref{Ireg} in odd dimensions can be divided into four
terms~\cite{Olea:2006vd, Miskovic:2008ck},
\begin{equation}
\delta I_{2n+1}^{\mathrm{KT}}=\delta I_{2n+1}^{(i)}+\delta
I_{2n+1}^{(ii)}+\delta I_{2n+1}^{(iii)}+\delta I_{2n+1}^{(iv)}\,.
\label{i-iv}
\end{equation} The main idea is to re-arrange the different terms in such a way that holographic expansions of the corresponding fields can be implemented. With this at hand, and armed with a power-counting procedure, one may recognize the finite part of the variation of the action. This goes together with the action of Weyl transformations on the holographic metric and the subleading coefficients in the FG series, as they are expressible as covariant functions of $g_{(0)ij}$~\cite{Imbimbo:1999bj}.

The first term comes from the variation of the boundary Riemann tensor, that is,
\begin{eqnarray}
\delta I_{2n+1}^{(i)} &=& n(n-1)c_{2n}\int\limits_{\partial \mathcal{M}} d^{2n}x\,\sqrt{-h}\int\limits_{0}^{1}dt\int\limits_{0}^{t}ds\,\delta
_{i_{1}\cdots i_{2n}}^{j_{1}\cdots j_{2n}}\,K_{j_{1}}^{i_{1}}\delta
_{j_{2}}^{i_{2}}\,
h^{i_{4}k}\delta \mathcal{R}_{\ \ kj_{3}j_{4}}^{i_{3}} \\
&&\times \left( \frac{1}{2}\,\mathcal{R}_{j_{5}j_{6}}^{i_{5}i_{6}}-^{2}K_{j_{5}}^{i_{5}}K_{j_{6}}^{i_{6}}+\frac{s^{2}}{\ell ^{2}}\,\delta _{j_{5}}^{i_{5}}\delta_{j_{6}}^{i_{6}}\right)
\cdots \left( \frac{1}{2}\,\mathcal{R}%
_{j_{2n-1}j_{2n}}^{i_{2n-1}i_{2n}}-t^{2}K_{j_{2n-1}}^{i_{2n-1}}K_{j_{2n}}^{i_{2n}}+%
\frac{s^{2}}{\ell ^{2}}\,\delta _{j_{2n-1}}^{i_{2n-1}}\delta
_{j_{2n}}^{i_{2n}}\right) ,\notag
\end{eqnarray}
where, in turn, 
\begin{equation}
h^{i_{4}k}\delta \mathcal{R}_{\ \ kj_{3}j_{4}}^{i_{3}}=\frac{1}{2}\nabla
_{j_{3}}\left[ \nabla ^{i_{4}}\left( h^{-1}\delta h\right)_{j_{4}}^{i_{3}}-\nabla _{j_{4}}\delta h^{i_{3}i_{4}}-\nabla ^{i_{3}}\left( h^{-1}\delta h\right) _{j_{4}}^{i_{4}}
\right]-(j_3 \leftrightarrow j_4) \,. \label{dR}
\end{equation}
This expression originally involves the variation $\delta \mathcal{R}_{j_{3}j_{4}}^{i_{3}i_{4}}$, which has a piece in terms of the variation of the boundary metric 
\begin{equation} \label{deltaR}
\delta \mathcal{R}_{j_{3}j_{4}}^{i_{3}i_{4}}=\left( h^{-1}\delta h\right)_{k}^{i_{4}}\mathcal{R}_{j_{3}j_{4}}^{ki_{3}}+h^{i_{4}k}\delta \mathcal{R}_{\ \ kj_{3}j_{4}}^{i_{3}}\,.
\end{equation}
However, only the second term contributes to $\delta I_{2n+1}^{(i)}$. The first one, together with the variation of $\sqrt{-h}$, can be rewritten as $\delta I_{2n+1}^{(iii)}$ and a part of $\delta I_{2n+1}^{(ii)}$ (defined below) using an identity that follows from the vanishing of an antisymmetric Kronecker delta whose rank exceeds the boundary dimension. The derivation is presented in detail in Appendix \ref{AP:Variation}.

After all the manipulations mentioned above, the resulting contribution to the variation of the action is
\begin{eqnarray}
&&\delta I_{2n+1}^{(i)} = -2n(n-1)c_{2n}\int\limits_{\partial \mathcal{M}
} d^{2n}x\,\sqrt{-h}\int\limits_{0}^{1}dt\int\limits_{0}^{t}ds\,\delta_{i_{1} \cdots i_{2n}}^{j_{1}\cdots j_{2n}}\,K_{j_{1}}^{i_{1}}\delta_{j_{2}}^{i_{2}} \, \nabla_{j_{3}} \nabla ^{i_{3}}\left( h^{-1}\delta h\right) _{j_{4}}^{i_{4}}  \times \notag  \\
&& \left( \frac{1}{2}\,\mathcal{R}%
_{j_{5}j_{6}}^{i_{5}i_{6}}-t^{2}K_{j_{5}}^{i_{5}}K_{j_{6}}^{i_{6}}+\frac{
s^{2}}{\ell ^{2}}\,\delta _{j_{5}}^{i_{5}}\delta _{j_{6}}^{i_{6}}\right)
\cdots \left( \frac{1}{2}\,\mathcal{R}%
_{j_{2n-1}j_{2n}}^{i_{2n-1}i_{2n}}-t^{2}K_{j_{2n-1}}^{i_{2n-1}}K_{j_{2n}}^{i_{2n}}+%
\frac{s^{2}}{\ell ^{2}}\,\delta _{j_{2n-1}}^{i_{2n-1}}\delta_{j_{2n}}^{i_{2n}}\right) \,. \qquad
\label{var I(i)}
\end{eqnarray}
The second term in the variation of the total action is
\begin{eqnarray}
&&\delta I_{2n+1}^{(ii)} =\frac{1}{2^{n-1}}\int\limits_{\partial \mathcal{M}%
}d^{2n}x\,\sqrt{-h}\,\delta _{i_{1}\cdots i_{2n}}^{j_{1}\cdots j_{2n}}\,%
\left[ \left( h^{-1}\delta h\right) _{k}^{i_{1}}K_{j_{1}}^{k}+2\delta
K_{j_{1}}^{i_{1}}\right] \delta _{j_{2}}^{i_{2}} \times \label{varI(ii)} \\
&& \left[ \frac{2^{n-1}\,\delta
_{j_{3}}^{i_{3}}\cdots \delta _{j_{2n}}^{i_{2n}}}{16\pi G \,(2n-1)!}
+nc_{2n}\int\limits_{0}^{1}dt\,\left( R_{j_{3}j_{4}}^{i_{3}i_{4}}+
\frac{2t^{2}}{\ell ^{2}}\,\delta _{j_{3}}^{i_{3}}\delta
_{j_{4}}^{i_{4}}\right) \cdots \left( R_{j_{2n-1}j_{2n}}^{i_{2n-1}i_{2n}}+%
\frac{2t^{2}}{\ell ^{2}}\,\delta _{j_{2n-1}}^{i_{2n-1}}\delta
_{j_{2n}}^{i_{2n}}\right) \right] \,. \notag
\end{eqnarray}
In Einstein spaces, the spacetime Weyl tensor takes the simpler form
\begin{equation}
    W^{\alpha \beta}_{\mu \nu}=R^{\alpha \beta}_{\mu \nu} + \dfrac{1}{\ell^2}\, \delta^{\alpha \beta}_{\mu \nu} \, .  \label{BulkW}
\end{equation}
This is important to write down the formula~\eqref{varI(ii)} in terms of a polynomial of the bulk  Weyl tensor with boundary indices, which adopts the form
\begin{eqnarray}
&&\delta I_{2n+1}^{(ii)} =\frac{1}{2^{n-1}}\int\limits_{\partial
\mathcal{M}}d^{2n}x\,\sqrt{-h}\,\delta _{i_{1}\cdots i_{2n}}^{j_{1}\cdots
j_{2n}}\,\left[ \left( h^{-1}\delta h\right)
_{k}^{i_{1}}K_{j_{1}}^{k}+2\delta K_{j_{1}}^{i_{1}}\right] \delta_{j_{2}}^{i_{2}}\left[ \frac{2^{n-1}\,\delta _{j_{3}}^{i_{3}}\cdots \delta
_{j_{2n}}^{i_{2n}}}{16\pi G \,(2n-1)!}+\right.   \notag \\
&&+\left. nc_{2n}\int\limits_{0}^{1}dt\,\left( W_{j_{3}j_{4}}^{i_{3}i_{4}}+
\frac{t^{2}-1}{\ell ^{2}}\,\delta_{j_{3}j_4}^{i_{3} i_4}\right) \cdots \left( W_{j_{2n-1}j_{2n}}^{i_{2n-1}i_{2n}}+
\frac{t^{2}-1}{\ell ^{2}}\,\delta _{j_{2n-1}j_{2n}}^{i_{2n-1}i_{2n}}\right) \right] . \label{(ii)Wpolynomial}
\end{eqnarray}
A remarkable property of $\delta I_{2n+1}^{(ii)}$ is that it can be factorized by the
Weyl tensor. In order to see this, one recalls the fact that the coefficient $c_{2n}$  comes from a double parametric integration in Eq.~\eqref{c2n}. In doing so, this part of the  variation can be schematically rewritten as
\begin{eqnarray}
\delta I_{2n+1}^{(ii)} &=&\frac{nc_{2n}}{2^{n-1}}\int\limits_{\partial
\mathcal{M}}d^{2n}x\,\sqrt{-h}\,\delta ^{[2n]}\,\delta \left[
\rule[2pt]{0pt}{11pt}\left( h^{-1}\delta h\right) \cdot K+2\delta K\right]\times
\notag \\
&&\times \int\limits_{0}^{1}dt\,\left[ \left( W+\frac{t^{2}-1}{\ell ^{2}}\,\delta
^{[2]}\right) ^{n-1}-\left( \frac{t^{2}-1}{\ell ^{2}}\,\delta
^{[2]}\right) ^{n-1}\right].
\end{eqnarray}
The rest of the variation considers one term which goes along $h^{-1}\delta h$,
\begin{eqnarray} \label{(iii)}
&&\delta I_{2n+1}^{(iii)} =-nc_{2n}\int\limits_{\partial
\mathcal{M}}d^{2n}x\,\sqrt{-h}\,\int\limits_{0}^{1}dt\,t\,\delta_{i_{1}\cdots i_{2n}}^{j_{1}\cdots j_{2n}}\!\left( h^{-1}\delta h\right)
_{k}^{i_{1}}\left( K_{j_{1}}^{k}\delta _{j_{2}}^{i_{2}}-\delta
_{j_{1}}^{k}K_{j_{2}}^{i_{2}}\right) \times  \\
&&\times \left( \frac{1}{2}\mathcal{R}%
_{j_{3}j_{4}}^{i_{3}i_{4}}-t^{2}K_{j_{3}}^{i_{3}}K_{j_{4}}^{i_{4}}+\frac{t^{2}}{\ell ^{2}}\,\delta_{j_{3}}^{i_{3}}\delta _{j_{4}}^{i_{4}}\right)
\cdots \left(  \frac{1}{2}\mathcal{R}%
_{j_{2n-1}j_{2n}}^{i_{2n-1}i_{2n}}-t^{2}K_{j_{2n-1}}^{i_{2n-1}}K_{j_{2n}}^{i_{2n}}+%
\frac{t^{2}}{\ell ^{2}}\,\delta _{j_{2n-1}}^{i_{2n-1}}\delta
_{j_{2n}}^{i_{2n}}\right) \,, \notag 
\end{eqnarray}
and another along $\delta K$,
\begin{eqnarray}
\delta I_{2n+1}^{(iv)} &=&-2nc_{2n}\int\limits_{\partial
\mathcal{M}}d^{2n}x\,\sqrt{-h}\,\int\limits_{0}^{1}dt\,t\,\delta
_{i_{1}\cdots i_{2n}}^{j_{1}\cdots j_{2n}}\delta _{j_{1}}^{i_{1}}\delta
K_{j_{2}}^{i_{2}}\, \left( \frac{1}{2}\mathcal{R}%
_{j_{3}j_{4}}^{i_{3}i_{4}}-t^{2}K_{j_{3}}^{i_{3}}K_{j_{4}}^{i_{4}}+\frac{t^{2}}{\ell ^{2}}\,\delta_{j_{3}}^{i_{3}}\delta _{j_{4}}^{i_{4}}\right)
\times   \notag \\
&&\qquad \qquad \qquad \cdots \times \left(  \frac{1}{2}\mathcal{R}%
_{j_{2n-1}j_{2n}}^{i_{2n-1}i_{2n}}-t^{2}K_{j_{2n-1}}^{i_{2n-1}}K_{j_{2n}}^{i_{2n}}+%
\frac{t^{2}}{\ell ^{2}}\,\delta _{j_{2n-1}}^{i_{2n-1}}\delta
_{j_{2n}}^{i_{2n}}\right) \,.
\end{eqnarray}
In order to extract holographic information on the Weyl anomaly, it is important to identify the finite contribution from the generic variation of the action~\eqref{i-iv}, once a Weyl transformation has been performed in the asymptotic form of the fields. This asymptotic analysis is carried out in detail in the next section.

\section{Near-boundary analysis in Einstein-AdS gravity }

The holographic reconstruction of AAdS spacetimes is given by the determination of the coefficients in FG expansion in terms of the holographic data.
For a generic metric at the conformal boundary in Eq.~\eqref{FGexpansion}, the expansion of its inverse reads
\begin{equation}
\bar{g}^{ij}=g_{(0)}^{ij}-\frac{z^2}{\ell^2}\,g_{(2)}^{ij}+\frac{z^{4}}{\ell^4}\,\left(
g_{(2)}^{ik}g_{(2)k}^{j}-g_{(4)}^{ij}\right) +\mathcal{O}(z^{6})\,.
\end{equation}
On the other hand, the square root of the determinant of $h_{ij}$ dictates the highest-order divergences due to the fact that
\begin{equation}
\sqrt{-h}=\frac{\ell^{2n}}{z^{2n}}\,\sqrt{-\bar{g}}\,,
\label{det_h}
\end{equation}
where
\begin{equation}
\sqrt{-\bar{g}}=\sqrt{-g_{(0)}}\left[ 1-\frac{z^{2}}{2}\,\mathcal{S}_{(0)}+\frac{z^{4}}{8}\,\left( \mathcal{S}_{(0)}^{2}-\mathcal{S}_{(0)ij}\mathcal{S}_{(0)}^{ij}\right) \right] +
\mathcal{O}(z^{6})\,.
\label{vol}
\end{equation}
Taking into account the definition of the extrinsic curvature given by~\eqref{Kij}, its near-boundary form is expressed as
\begin{equation}
K_{j}^{i}=\frac{1}{\ell }\,\delta _{j}^{i}+\frac{z^{2}}{\ell }\, \mathcal{S}_{(0)j}^{i} +\frac{z^4}{2\ell }\,\left(\frac{1}{d-4}\, \mathcal{B}^i_{(0)j}+\mathcal{S}^i_{(0)k}\mathcal{S}^k_{(0)j}\right)+\mathcal{O}(z^6)\,.
\label{K.expanded}
\end{equation}
By definition, curvatures computed for a given boundary metric do not contain normal derivatives, such that $z$ can be regarded as constant. Then the rescaling in the metric $h_{ij}=(\ell^2/z^2)\bar{g}_{ij}$ produces a rescaling of the Riemann tensor
\begin{equation}
\mathcal{R}_{kl}^{ij}\left( h\right) =\frac{z^{2}}{\ell^2}\,\mathcal{R}_{kl}^{ij}\left(
\bar{g}\right) \,.
\end{equation}
In turn, the Riemann tensor for the metric at the conformal boundary expands as
\begin{equation}
\mathcal{R}_{kl}^{ij}(\bar{g})=\mathcal{R}_{(0)kl}^{ij}+{z}^{2}\left(
\mathcal{R}_{(0)m[k}^{ij}\mathcal{S}_{(0)l]}^{m}-2\mathcal{R}_{(0)kl}^{m[i}
\mathcal{S}_{(0)m}^{j]}-\nabla ^{[i}\mathcal{C}_{(0)kl}^{j]}\right) +
\mathcal{O}(z^{4})\,, 
\end{equation}
what, combined with the Gauss-Codazzi relations (see Appendix~\ref{GC}), allows one to determine the independent components of the bulk Weyl tensor.

In particular, the electric part of the Weyl tensor considers the  contraction of the Weyl tensor with two normal vectors to the boundary, such that in Gauss-normal frame, reads
\begin{equation}
\mathcal{E}^{i}_{j}=\frac{1}{(d-2)} \, W^{i\mu}_{j  \nu}\, n_{\mu}n^{\nu}=\frac{1}{(d-2)} \, W^{iz}_{j z}\,  .  \label{ElecW}
\end{equation}
Furthermore, the fact the Weyl tensor is traceless implies that the electric part of the Weyl tensor can be also written as
\begin{equation}
W^{iz}_{j z}=-W_{kj}^{ki}\,. \label{E-W}
\end{equation}
This part of the Weyl tensor behaves asymptotically as
\begin{equation}
W^{iz}_{jz} =  \frac{z^4}{\left(d-4\right) \ell^2 } \, \mathcal{B}^i_{\left(0\right)j}+ \mathcal{O}(z^6) \,,
    \label{electric}
\end{equation}
while the other components can be readily expanded as
\begin{align}
    W_{kl}^{ij} &= \frac{z^{2}}{\ell ^{2}}\,\mathcal{W}_{\left(0\right)kl}^{ij}
-\frac{z^4}{\ell^2}\left[\frac{2}{d-4}\,\mathcal{B}_{\left(0\right)[k}^{[i}\delta
_{l]}^{j]}+\left( 2\mathcal{S}_{\left(0\right)m}^{[i}\mathcal{W}_{\left(0\right)kl}^{j]m}-\mathcal{S}_{\left(0\right)[k}^{m}\mathcal{W}_{\left(0\right)l]m}^{ij}\right)+\,\nabla
^{[i}\mathcal{C}^{j]}_{\left(0\right)kl}\right]+\mathcal{O}(z^6)\,,
\notag\\
W^{i z}_{j k} &= - \frac{ z^3}{\ell^2}  \mathcal{C}_{\left(0\right)j k}^{i} + \mathcal{O} \left(z^5\right) \,.
\label{magnetic}
\end{align}
Notice that, in ACF solutions of AdS gravity, all components of the Weyl tensor fall off at normalizable order, i.e., are $\mathcal{O}(z^d)$. In this case, no divergences in the holographic coordinate $z$ appear in the action~\eqref{Ireg}, up to the logarithmic term, as shown in Ref.~\cite{Anastasiou:2020zwc}.  

However, this is no longer true for a generic AAdS space.

\section{Weyl transformations and Weyl anomaly}

The Penrose-Brown-Henneaux transformations are a subset of diffeomorphisms that leave invariant the form of the Fefferman-Graham metric~\cite{Imbimbo:1999bj}. In particular, the relation to conformal structures at the boundary is given by the fact that radial diffeomorphisms induce Weyl transformations for the metric $g_{(0)ij}$. For a Weyl transformation expressed in terms of the parameter $\sigma (x)$, the holographic metric changes as 
\begin{equation}
\delta _{\sigma }g_{(0)ij} =2\sigma \,g_{(0)ij}\,.  \label{Weylg0}
\end{equation}
Using standard holographic techniques for the computation of boundary correlators, this Weyl rescaling leads to the corresponding conformal anomaly. 
In contrast, in the present approach, one also needs the transformation law for Weyl rescalings for higher-order coefficients, produced by the relation~\eqref{Weylg0}. These expressions are given by
\begin{eqnarray}
\delta _{\sigma }g_{(2)ij} &=&\frac{\ell ^{2}}{2}\,\left( \nabla
_{(0)i}\nabla _{(0)j}\sigma +\nabla _{(0)j}\nabla _{(0)i}\sigma \right) \,,
\notag \\
\delta _{\sigma }g_{(4)ij} &=&-2\sigma \,g_{(4)ij}+\frac{\ell ^{2}}{2}%
\,\left( \nabla _{(0)}^{k}\sigma \nabla _{(0)k}g_{(2)ij}+g_{(2)i}^{k}\nabla
_{(0)j}\nabla _{(0)l}\sigma +g_{(2)j}^{k}\nabla _{(0)i}\nabla _{(0)k}\sigma
\right)   \notag \\
&&-\frac{\ell ^{2}}{4}\,\left( \nabla _{(0)i}g_{(2)j}^{k}\nabla
_{(0)k}\sigma +\nabla _{(0)j}g_{(2)i}^{k}\nabla _{(0)k}\sigma \right) \,.
\end{eqnarray}
When the FG modes are written in terms of curvatures associated to the holographic data, the above relations imply standard Weyl transformations of the boundary tensors~\cite{Osborn:2015rna}. In particular, the Schouten tensor transforms as
\begin{equation}
\delta _{\sigma }\mathcal{S}_{(0)j}^{i}=-2\sigma \,\mathcal{S}_{(0)j}^{i}-\frac{1}{2}\,\left( \nabla
_{(0)}^{i}\nabla _{(0)j}\sigma +\nabla _{(0)j}\nabla _{(0)}^{i}\sigma
\right) \,. \label{lawS}
\end{equation}
In a similar fashion, the Weyl rescaling of the metric leads to the Weyl transformation of the boundary Bach tensor, which is given by 
\begin{equation}
    \delta_\sigma \mathcal{B}_{(0)ij}=-2\sigma \mathcal{B}_{(0)ij} +(d-4)  (\mathcal{C}_{(0)ijk}+\mathcal{C}_{(0)jik})\nabla_{(0)}^k\sigma\,.
\end{equation}
Other relations that will be useful for the computation of holographic conformal anomalies are
\begin{equation}
\delta _{\sigma }\Gamma _{(0)jk}^{i}=\delta^{i}_{k}\nabla_{(0)j}\sigma+\delta^{i}_{j}\nabla_{(0)k}\sigma-g_{(0)jk}\nabla^{i}_{(0)}\sigma\,,
\end{equation}
\begin{align}
    \delta _{\sigma }\mathcal{R}_{(0)kl}^{ij}
    &=-2\sigma \mathcal{R}_{(0)kl}^{ij}-2\nabla^{[i}_{(0)}\nabla_{(0)[k}\sigma\delta^{j]}_{l]}-2\nabla_{(0)[k}\nabla^{[i}_{(0)}\delta^{j]}_{l]}\,.
\end{align}
As for Weyl variations acting on the boundary metric, it is relevant to render explicit the expansion of the quantity
\begin{eqnarray}
\left( h^{-1}\delta _{\sigma }h\right) _{k}^{i}K^{k}_{j} &=&\left( \bar{g}^{-1}\delta
_{\sigma }\bar{g}\right) _{k}^{i} K^{k}_{j} \notag \\
&=&2\sigma \,\left[
\delta _{j}^{i}+2z^{2}\mathcal{S}_{(0)j}^{i}+{z^{4}}\left( 2\mathcal{S}_{(0)k}^{i}\mathcal{S}_{(0)j}^{k}+
\frac{1}{(d-4)}\, \mathcal{B}_{(0)j}^{i}\right) +\mathcal{O}(z^{6})\right] . \quad 
\label{gdg}
\end{eqnarray}
Due to the fact that the expansion of the intrinsic curvature $K_{j}^{i}$ starts with a constant term, the homogeneous part of the Weyl transformation is subleading with respect to the above formula
\begin{equation}
 \delta _{\sigma } K_{j}^{i} =      - 2\sigma \left[\frac{z^{2}}{\ell }\,\mathcal{S}^{i}_{(0)j}+\frac{z^{4}}{\ell}\left( \mathcal{S}_{(0)k}^{i}\mathcal{S}_{(0)j}^{k}+
\frac{1}{(d-4)}\, \mathcal{B}_{(0)j}^{i}\right) +\mathcal{O}(z^{6})\right],
\label{dK}
\end{equation}
what, in turn, implies
\begin{equation}
\left( h^{-1}\delta_\sigma h\right) _{k}^{i} K_{j}^{k}+2\delta_\sigma K_{j}^{i}=  2\sigma \,\left[ \frac{1}{\ell }\,\delta _{j}^{i}-\frac{z^{4}}{\ell (d-4)}\, \mathcal{B}_{(0)j}^{i}\right] +\mathcal{O}(z^{6})\,.
\label{dKcal}
\end{equation}
In what follows, the finite contributions in the variation of the total action are shown to give rise to the Weyl anomaly. Most of the derivation relies on a power-counting argument for the coordinate $z$. Parts (iii) and (iv) of the variation of the action are quite simple to render explicit, producing closed expressions in any odd dimension. In contrast, parts (i) and (ii) are far more involved and need to be worked out on a case-by-case basis. Below, they are presented in order of difficulty.

\subsection{Part (iii): vanishing contributions to anomalies}
\label{Sub: Vanishing(iii)}

The simplest part of the Weyl variation of the action is  the one in Eq.~\eqref{(iii)}
\begin{align} \label{(iii-Weyl)}
\delta_\sigma I_{2n+1}^{(iii)} &=-nc_{2n}\int\limits_{\partial
\mathcal{M}}d^{2n}x\,\sqrt{-h}\,\int\limits_{0}^{1}dt\,t\,\delta_{i_{1}\cdots i_{2n}}^{j_{1}\cdots j_{2n}}\!\left( h^{-1}\delta_\sigma h\right)
_{k}^{i_{1}}\left( K_{j_{1}}^{k}\delta _{j_{2}}^{i_{2}}-\delta
_{j_{1}}^{k}K_{j_{2}}^{i_{2}}\right) \times  \\
& \left( \frac{1}{2}\mathcal{R}%
_{j_{3}j_{4}}^{i_{3}i_{4}}-t^{2}K_{j_{3}}^{i_{3}}K_{j_{4}}^{i_{4}}+\frac{t^{2}}{\ell ^{2}}\,\delta_{j_{3}}^{i_{3}}\delta _{j_{4}}^{i_{4}}\right) 
\cdots \left(  \frac{1}{2}\mathcal{R}%
_{j_{2n-1}j_{2n}}^{i_{2n-1}i_{2n}}-t^{2}K_{j_{2n-1}}^{i_{2n-1}}K_{j_{2n}}^{i_{2n}}+%
\frac{t^{2}}{\ell ^{2}}\,\delta _{j_{2n-1}}^{i_{2n-1}}\delta
_{j_{2n}}^{i_{2n}}\right) \,. \notag
\end{align}
Indeed, one does not need further information beyond the leading-order terms in the relations
\begin{eqnarray}
    \sqrt{-h}&=&\frac{\ell^{2n}}{z^{2n}}\sqrt{-g_{0}}+\cdots\,, \label{expandsqrt} \\
 \left( h^{-1}\delta h \right)_{k}^{i} &=& \left( g_{(0)}^{-1}\delta g_{(0)}\right)_{k}^{i}+\cdots\,,
\end{eqnarray}
in order to prove that this expression is, at most, finite.

One can also find the expansion of the different pieces that form the expression~\eqref{(iii)} by simple inspection of the asymptotic form of the extrinsic curvature~\eqref{K.expanded}, that is,
\begin{eqnarray}
\left( K_{j_{1}}^{k}\delta _{j_{2}}^{i_{2}}-\delta
_{j_{1}}^{k}K_{j_{2}}^{i_{2}}\right) &=&\ell z^2 \left( \mathcal{S}_{(0)j_{1}}^{k}\delta _{j_{2}}^{i_{2}}-\delta_{j_{1}}^{k} \mathcal{S}_{(0)j_{2}}^{i_{2}}\right) +\mathcal{O}(z^4)\,,  \label{expbracket} \\
\left( \frac{1}{2}\mathcal{R}_{j_{3}j_{4}}^{i_{3}i_{4}}(h) -t^{2}K_{j_{3}}^{i_{3}}K_{j_{4}}^{i_{4}}+\frac{t^{2}}{\ell ^{2}}\,\delta _{j_{3}}^{i_{3}}\delta _{j_{4}}^{i_{4}}\right) &=&\frac{z^2}{\ell^2}\left(\frac{1}{2}\mathcal{R}_{(0)j_{3}j_{4}}^{i_{3}i_{4}}-t^{2}\left( \delta
_{j_{3}}^{i_{3}}\mathcal{S}_{(0)j_{4}}^{i_{4}}+\mathcal{S}_{(0)j_{3}}^{i_{3}}\,
\delta _{j_{4}}^{i_{4}}\right)\right) +\mathcal{O}(z^4). \notag
\end{eqnarray}
Upon a trivial change of variable in the integration $u=t^2$, and plugging in  all the above results in Eq.~\eqref{(iii)}, there is a  finite part
\begin{eqnarray}
\delta_\sigma I_{2n+1}^{(iii)} &=&-\frac{nc_{2n}\ell }{2^{n}}\int\limits_{\partial
\mathcal{M}}d^{2n}x\,\sqrt{-g_{(0)}}\int\limits_{0}^{1}du\,\delta
_{i_{1}\cdots i_{2n}}^{j_{1}\cdots j_{2n}}\!\left( g_{(0)}^{-1}\delta_\sigma
g_{(0)}\right) _{k}^{i_{1}}\left( \mathcal{S}_{(0)j_{1}}^{k}\delta
_{j_{2}}^{i_{2}}-\delta _{j_{1}}^{k}\mathcal{S}_{(0)j_{2}}^{i_{2}}\right)
\times \notag  \\
&&\times \left( \mathcal{R}_{(0)j_{3}j_{4}}^{i_{3}i_{4}}-4u\,\delta
_{j_{3}}^{i_{3}}\mathcal{S}_{(0)j_{4}}^{i_{4}}\right) \cdots \left( \mathcal{R}%
_{(0)j_{2n-1}j_{2n}}^{i_{2n-1}i_{2n}}-4u\,\delta _{j_{2n-1}}^{i_{2n-1}}
\mathcal{S}_{(0)j_{2n}}^{i_{2n}}\right) \,,  
\end{eqnarray}
plus terms with positive powers of $z$.

The corresponding Weyl transformation on the holographic metric produces
\begin{equation}
\left( g_{(0)}^{-1}\delta_{\sigma}
 g_{(0)}\right) _{k}^{i_{1}}
=2\sigma \,\delta^{i_{1}}_k\,,
\end{equation}%
what directly implies
\begin{equation}
\delta _{\sigma} I_{2n+1}^{(iii)}=0\,.
\end{equation}

\subsection{Part (iv): type A anomaly and Pfaffian of the Weyl tensor}
\label{Sub: (iv)}

In a similar line of reasoning as in the previous subsection, the type A anomaly can be read off from the finite part of $\delta_{\sigma}I_{2n+1}^{(iv)}$. The Pfaffian of $\mathcal{W}_{(0)}$, which is the totally antisymmetric product of $n$ Weyl tensor, and part of the type B anomaly, can also be obtained at the boundary of any odd-dimensional AAdS spacetime.

From the generic formula for the variation~\eqref{i-iv}, that part can be written as
\begin{eqnarray}
\delta_\sigma I_{2n+1}^{(iv)} &=&-2nc_{2n}\int\limits_{\partial
\mathcal{M}}d^{2n}x\,\sqrt{-h}\,\int\limits_{0}^{1}dt\,t\,\delta
_{i_{1}\cdots i_{2n}}^{j_{1}\cdots j_{2n}}\delta _{j_{1}}^{i_{1}}\delta_\sigma
K_{j_{2}}^{i_{2}}\, \left( \frac{1}{2}\mathcal{R}%
_{j_{3}j_{4}}^{i_{3}i_{4}}-t^{2}K_{j_{3}}^{i_{3}}K_{j_{4}}^{i_{4}}+\frac{t^{2}}{\ell ^{2}}\,\delta_{j_{3}}^{i_{3}}\delta _{j_{4}}^{i_{4}}\right)
\times   \notag \\
&&\qquad \qquad \qquad \cdots \times \left(  \frac{1}{2}\mathcal{R}%
_{j_{2n-1}j_{2n}}^{i_{2n-1}i_{2n}}-t^{2}K_{j_{2n-1}}^{i_{2n-1}}K_{j_{2n}}^{i_{2n}}+%
\frac{t^{2}}{\ell ^{2}}\,\delta _{j_{2n-1}}^{i_{2n-1}}\delta
_{j_{2n}}^{i_{2n}}\right) \,.
\end{eqnarray}
The proper use of the relations~\eqref{expandsqrt} and~\eqref{expbracket}, a change of variable in the integral, together with the leading order of~\eqref{dK}, produces a finite contribution of the form
\begin{eqnarray}
\delta_\sigma I_{2n+1}^{(iv)} &=&-\frac{\ell nc_{2n} }{2^{n-1}}\int\limits_{\partial
\mathcal{M}}d^{2n}x\,\sqrt{-g_{(0)}}\int\limits_{0}^{1}du\,\delta
_{i_{1}\cdots i_{2n}}^{j_{1}\cdots j_{2n}}\,\delta
_{j_1}^{i_{1}}\delta_\sigma \mathcal{S}_{(0)j_2}^{i_2}
\times \notag  \\
&&\times \left( \mathcal{R}_{(0)j_{3}j_{4}}^{i_{3}i_{4}}-4u\,\delta
_{j_{3}}^{i_{3}}\mathcal{S}_{(0)j_{4}}^{i_{4}}\right) \cdots \left( \mathcal{R}%
_{(0)j_{2n-1}j_{2n}}^{i_{2n-1}i_{2n}}-4u\,\delta _{j_{2n-1}}^{i_{2n-1}}
\mathcal{S}_{(0)j_{2n}}^{i_{2n}}\right) \,.  \label{IV-Schouten}
\end{eqnarray}
For the rest of the derivation, one can proceed schematically due to the fact that the above expression involves tensors with equal number of indices up than down. This simply means one can freely move any object around, as long as the indices structure is preserved. For simplicity, one can take the notation $\delta _{j}^{i} \to \delta $, $\delta _{i_{1}\cdots i_{2n}}^{j_{1}\cdots j_{2n}} \to \delta ^{[2n]}$, $K_{j}^{i} \to K$, $\mathcal{R}^{i_1i_2}_{j_1j_2}\to \mathcal{R}$, etc. Employing this compact notation, the last relation can be cast as
\begin{equation}
\delta _{\sigma} I_{2n+1}^{(iv)}=-\frac{\ell nc_{2n}}{2^{n-1}}
\int\limits_{\partial \mathcal{M}}d^{2n}x\,\sqrt{-g_{(0)}}\,\delta ^{[
2n]}\delta \,\delta _{\sigma }\mathcal{S}_{(0)}
\int\limits_{0}^{1}du\,\!\left( \rule[2pt]{0pt}{11pt}\mathcal{R}_{(0)}-4u\delta \mathcal{S}_{(0)} \right) ^{n-1}\,.
\end{equation}
The homogeneous part of the Weyl transformation~\eqref{lawS} produces a polynomial expression which is easy to manipulate, regardless the particular value of $n$. In fact, this term is conveniently written as
\begin{equation}
\delta _{\sigma }I_{2n+1}^{(iv)}=- \frac{\ell nc_{2n}}{2^{n-2}}\int\limits_{\partial \mathcal{M}}d^{2n}x\,\sqrt{-g_{(0)}}\,\sigma\,\delta
^{[ 2n]}\delta \,\mathcal{S}_{(0)}
\int\limits_{0}^{1}du\,\left( \rule[2pt]{0pt}{11pt}\mathcal{R}
_{(0)}-4u\delta \mathcal{S}_{(0)} \right) ^{n-1}\,,
\end{equation}
such that the identity for the binomial 
\begin{equation}\label{pbinom}
\left( a+b\right) ^{n}-a^{n}=nb\int\limits_{0}^{1}du\,\left( a+ub\right)
^{n-1}\,,
\end{equation}%
can be implemented upon the suitable substitution $a\rightarrow \mathcal{R}_{(0)}$ and $b\rightarrow -4\delta \mathcal{S}_{( 0) } $, what leads to
\begin{equation}
\delta _{\sigma } I_{2n+1}^{(iv)}=-\frac{\ell c_{2n}}{2^n}
\int\limits_{\partial \mathcal{M}}d^{2n}x\,\sqrt{-g_{(0)}}\,\sigma \,\delta ^{[2n]}
\left[ \left( \rule[2pt]{0pt}{11pt}\mathcal{R}_{(0)}-4\delta \mathcal{S}_{( 0) }\right) ^{n}-\mathcal{R}_{(0)}^{n}\right] \,.
\end{equation}
The antisymmetry of the indices in the Kroenecker delta, together with the expression of  the Weyl tensor~\eqref{W},  are such that the last formula boils down to the difference between the Pfaffian of the Weyl tensor and  the Euler density
\begin{equation}
\delta _{\sigma }I_{2n+1}^{(iv)}=-\ell c_{2n}
\int\limits_{\partial \mathcal{M}}d^{2n}x\,\sqrt{-g_{(0)}}\,\sigma \left( \rule[2pt]{0pt}{11pt} \mathrm{Pf}(\mathcal{W}_{(0)})-\mathcal{E}_{2n}\right) \,.
\end{equation}
Therefore, part (iv) of the Weyl variation of the total action contributes to this holographic computation with the type A  anomaly, and a particular combination of the maximal power of the Weyl tensor, i.e., its Pfaffian, to the type B anomaly,
\begin{equation}
\delta _{\sigma } I^{(iv)}_{2n+1}= \int\limits_{\partial \mathcal{M}
}d^{2n}x\,\sqrt{-g_{(0)}}\,\sigma \,(\mathcal{A}_{\mathrm{A}}+\mathcal{A}_{\mathrm{B}}^{(iv)})\,,
\label{TypeA-PfB}
\end{equation}
where 
\begin{eqnarray}
\mathcal{A}_{\mathrm{A}} &=&
\ell c_{2n}\mathcal{E}_{2n}
\,,\notag\\
\mathcal{A}_{\mathrm{B}}^{(iv)}&=&
 -\ell c_{2n}\, \mathrm{Pf}(\mathcal{W}_{(0)})\,.
\end{eqnarray}
The corresponding central charges $a$ and $c$ are easily read off as
\begin{equation}
a=c=(-1)^{n}\frac{1}{16\pi G}\frac{n\ell ^{2n-1}}{2^{2n-2}n!^{2}}\,,
\label{a_2n}
\end{equation}
in agreement with Ref.~\cite{Imbimbo:1999bj}.

In general, a Weyl tranformation acting on a field may induce an inhomogeneous part which involves covariant derivatives of the conformal factor $\sigma$. This is the case of the Schouten tensor $\mathcal{S}^i_{(0)j}$ in Eq.~\eqref{IV-Schouten}. However, as long as the corresponding part of the variation  is both finite and expressible in terms of covariant tensors of $g_{(0)}$, one can always integrate by parts with respect to the covariant derivative $\nabla _{(0)i}$. Indeed, assuming that the boundary is smooth ($\partial \partial \mathcal{M}=0$), this inhomogeneous piece is guaranteed to adopt the form
\begin{equation}
\left(\delta _{\sigma }I_{2n+1}\right)_{\mathrm{inhom}}=\int\limits_{\partial \mathcal{M}%
}d^{2n}x\,\sqrt{-g_{\left(0\right)}}\,\nabla _{\left(0\right)i}\sigma
\,V^{i}(g_{\left(0\right)})=\int\limits_{\partial \mathcal{M}}d^{2n}x\,\sqrt{-g_{\left(0\right)}}\,\sigma
\,\nabla _{\left(0\right)i}V^{i}\left(g_{\left(0\right)}\right)\,,
\end{equation}
when dropping boundary terms. 

Therefore, any inhomogeneous contribution to the Weyl variation of the total action, at the finite holographic order, would be a part of the type C anomaly,
\begin{equation}
(\delta _{\sigma }I_{2n+1})_{\mathrm{inhom}}=\int\limits_{\partial \mathcal{M}}d^{2n}x\,\sqrt{-g_{(0)}}\,\sigma \,\mathcal{A}_{\mathrm{C}}\,,
\end{equation}
whose explicit form will not be provided here.

\subsection{Part (ii): contribution to type B anomaly}
\label{Sub: Type B easy}

On the contrary to parts (iii) and (iv), which can be worked out in an arbitrary even boundary dimension, only an algorithm which leads to the contribution to the holographic anomaly can be outlined for parts (i) and (ii): no closed expression can be obtained in general. The obvious reason for that is the fact these contributions are not manifestly finite, but also contain divergent terms, as a consequence of the mismatch between Kounterterms prescription and the standard Holographic Renormalization procedure.

The fact that Eq.~\eqref{(ii)Wpolynomial} is a polynomial for the Weyl tensor and Kronecker delta $\delta^{[2]}$, that vanishes for global AdS spacetime, suggests a factorization by $W$.  As a matter of fact, because this part of the variation of the action gives rise to the mass of a black hole in AdS gravity~\cite{Olea:2006vd}, it is indeed controlled by the asymptotic behavior of the Weyl tensor, as it can be cast in the form
\begin{eqnarray}
&&\delta I_{2n+1}^{(ii)} =\frac{n(n-1)c_{2n}}{2^{n-1}}\int\limits_{\partial
\mathcal{M}}d^{2n}x\,\sqrt{-h}\,\delta _{i_{1}\cdots i_{2n}}^{j_{1}\cdots
j_{2n}}\,\left[ \left( h^{-1}\delta h\right)
_{k}^{i_{1}}K_{j_{1}}^{k}+2\delta K_{j_{1}}^{i_{1}}\right] \delta_{j_{2}}^{i_{2}}\,
W_{j_{3}j_{4}}^{i_{3}i_{4}}\times   \notag \\
&&\times \int\limits_{0}^{1}dt\int\limits_{0}^{1}du\,\left(
\frac{t^{2}-1}{\ell ^{2}}\,\delta_{j_{5}j_{6}}^{i_{5}i_{6}}+uW_{j_{5}j_{6}}^{i_{5}i_{6}}\right)\cdots\left(
\frac{t^{2}-1}{\ell ^{2}}\,\delta_{j_{2n-1}j_{2n}}^{i_{2n-1}i_{2n}}+uW_{j_{2n-1}j_{2n}}^{i_{2n-1}i_{2n}}\right)\,.
\label{I2weyl}\qquad 
\end{eqnarray}%
For ACF spaces in AdS gravity, W is of holographic order, i.e., $\mathcal{O}(z^{2n})$. On the other hand, $\sqrt{-h}\sim \mathcal{O}(z^{-2n})$, such that, in order to get a finite variation, it is enough to identify the finite contributions in the rest of the formula. This amounts to keeping the finite term of the square bracket in the first line 
\begin{equation}
\left[ \left( h^{-1}\delta h\right)_{k}^{i_{1}} K_{j_{1}}^{k}
+ 2\,\delta K_{j_{1}}^{i_{1}} \right] = \dfrac{1}{\ell}\left(g_{(0)}^{-1}\delta g_{(0)}\right)^{i_1}_{j_1}+\mathcal{O}(z^2)\,.
\end{equation}
Because there is no $\mathcal{O}(1)$ part in the Weyl tensor, the argument is equivalent to drop every single $W$ in the second line.

In doing so, one obtains
\begin{eqnarray}
\delta I_{2n+1}^{(ii)} &=&\frac{n(n-1)c_{2n}}{2^{n-1}}\int\limits_{\partial
\mathcal{M}}d^{2n}x\,\frac{\sqrt{-g_{(0)}}}{z^{2n}}\,\delta _{i_{1}\cdots i_{2n}}^{j_{1}\cdots
j_{2n}}\,\left[ \dfrac{1}{\ell}\left(g_{(0)}^{-1}\delta g_{(0)}\right)^{i_1}_{j_1}\right] \delta_{j_{2}}^{i_{2}}\,
W_{j_{3}j_{4}}^{i_{3}i_{4}}\times   \notag \\
&&\qquad \qquad \times \int\limits_{0}^{1}dt\left(
\frac{t^{2}-1}{\ell ^{2}}\,\right)^{n-2} \times 2^{n-2}\times
\delta^{i_5}_{j_5}\delta^{i_6}_{j_6} \dots \delta^{i_{2n-1}}_{j_{2n-1}}\delta^{i_{2n}}_{j_{2n}}\,,
\label{I2weylholord}
\end{eqnarray}%
that reflects the fact that variations of the extrinsic curvature do not contribute to the holographic, physical quantities of the theory. Without loss of generality, one may express part (ii) of the variation purely in terms of the variation of $h_{ij}$. Other contributions are, in general, subleading and vanish at the conformal boundary.

Upon the use of the explicit value for the coupling $c_{2n}$ and the integral
\begin{equation}
\int\limits_{0}^{1} dt\, (t^{2}-1)^{k}
= \frac{(-1)^{k} 2^{2k} (k!)^{2}}{(2k+1)!}\,,
\end{equation}
the result adopts the compact form
\begin{eqnarray}
\delta I_{2n+1}^{(ii)} &=&\frac{\ell^2}{64\pi G_{N} \left(2n-2\right)}\int\limits_{\partial
\mathcal{M}}d^{2n}x\,\sqrt{-h}\,\delta _{i_{1} i_{2}i_{3}}^{j_{1}
j_{2} j_{3}}\, \left( h^{-1}\delta h\right)
_{j_{1}}^{i_{1}}\,
W_{j_{2}j_{3}}^{i_{2}i_{3}} 
\,.
\end{eqnarray}%
Due to the fact that the double sub-trace in the Weyl tensor is zero, the identity  
\begin{equation}    
\delta^{j_{1} j_{2} j_{3}}_{i_{1} i_{2} i_{3}}
\, W_{j_{2} j_{3}}^{i_{2} i_{3}} = -4\, W_{i_{1} k}^{j_{1} k}\,,
\label{antisymmetricweylbound}
\end{equation}
can be combined with Eqs.~\eqref{E-W} and~\eqref{ElecW} to finally produce
\begin{equation}
\delta I_{2n+1}^{(ii)} =
\frac{\ell}{16\pi G_{N}} 
\int\limits_{\partial \mathcal{M}} d^{2n}x \,\sqrt{-h}\,
\mathcal{E}^{ij}\,\delta h_{ij},
\end{equation}
what matches the definition of Ashtekar--Magnon--Das charges
\cite{AshtekarMagnon1984, Ashtekar:1999jx, Jatkar:2014npa}. Needless to say, in this case, this part cannot contribute to the trace anomaly.

For a generic background metric for the CFT, it is more useful to take a step back to the formula
\begin{align}
&\delta I_{2n+1}^{(ii)} =\frac{1}{2^{n-1}} \int\limits_{\partial
\mathcal{M}}d^{2n}x\,\sqrt{-h}\,\delta _{i_{1}\cdots i_{2n}}^{j_{1}\cdots
j_{2n}}\,\left[ \left( h^{-1}\delta h\right)
_{k}^{i_{1}}K_{j_{1}}^{k}+2\delta K_{j_{1}}^{i_{1}}\right] \delta_{j_{2}}^{i_{2}} \left[ \frac{2^{n-1}}{16\pi G \,(2n-1)!} \delta
^{i_3}_{j_3} \cdots \delta
^{i_{2n}}_{j_{2n}} \notag \right.
\notag \\
& \left. 
+nc_{2n}\int\limits_{0}^{1}dt\,\left( W^{i_3 i_4}_{j_3 j_4}+\frac{(t^{2}-1)}{\ell ^{2}}\,\delta
^{i_3 i_4}_{j_3 j_4}\right) \cdots \left( W^{i_{2n-1} i_{2n}}_{j_{2n-1} j_{2n}}+\frac{(t^{2}-1)}{\ell ^{2}}\,\delta
^{i_{2n-1} i_{2n}}_{j_{2n-1} j_{2n}} \right)\right] \,, \label{Weyl(ii)}
\end{align}
and determine additional contributions to the type B anomaly.
This expression can be further simplified once the parametric integration is performed, and can be cast in the form
\begin{align}
\delta I_{2n+1}^{(ii)}&= -\frac{n! c_{2n}}{2^{n-1}} \int\limits_{\partial
\mathcal{M}}d^{2n}x\,\sqrt{-h} \sum_{p=1}^{n-1}\frac{\left( -1\right)
^{n-p}(n-1-p)! 2^{3\left(n-1-p\right)}}{p!\,\ell ^{2\left(n-1-p\right)}} \times \notag \\
& \times \,\delta^{j_1 \cdots j_{2p+1}}_{i_1 \cdots i_{2p+1}} \,\left[ \left( h^{-1}\delta h\right)
_{k}^{i_{1}}K_{j_{1}}^{k}+2\delta K_{j_{1}}^{i_{1}}\right] W^{i_2 i_3}_{j_2 j_3} \cdots W^{i_{2p} i_{2p+1}}_{j_{2p} j_{2p+1}} \,.
\label{Iiipartgeneric}
\end{align}
This is a rather formal expression, whose explicit form should be worked out dimension by dimension. However, there is yet another contribution to the type B anomaly which can be rendered manifest for any value of $n$. It considers a particular term in the above sum, with the product of $(n-1)$ Weyl tensors
\begin{equation}
\delta_{\sigma} I_{2n+1}^{(ii)} = \frac{n c_{2n}}{2^{n-1}}  \int\limits_{\partial
\mathcal{M}}d^{2n}x\,\sqrt{-h} \,\delta^{j_1 \cdots j_{2n-1}}_{i_1 \cdots i_{2n-1}} \,\left[ \left( h^{-1}\delta_{\sigma} h\right)
_{k}^{i_{1}}K_{j_{1}}^{k}+2\delta _{\sigma} K_{j_{1}}^{i_{1}}\right] W^{i_2 i_3}_{j_2 j_3} \cdots W^{i_{2n-2} i_{2n-1}}_{j_{2n-2} j_{2n-1}} +\cdots \,.
\end{equation} 
The expansions the fields given by Eqs.~\eqref{det_h},~\eqref{vol},~\eqref{magnetic} and~\eqref{dKcal}, are such that the above formula contains a divergent part given by the product of boundary Weyl tensor, which is the first mismatching term with respect to Holographic Renormalization. Keeping the $\mathcal{O}(1)$ term in the square bracket, the corresponding finite order is given either by $\mathcal{O}(z^{2-2n)}$ in Eqs.~\eqref{det_h},~\eqref{vol} or $\mathcal{O}(z^{4})$ in Eq.~\eqref{magnetic}, what produces
\begin{align}
&\delta_{\sigma} I_{2n+1}^{(ii)} = -\frac{\sigma \ell n c_{2n}}{2^{n-2}}  \int\limits_{\partial
\mathcal{M}}d^{2n}x\,\sqrt{-g_{\left(0\right)}} \,\delta^{j_1 \cdots j_{2n-2}}_{i_1 \cdots i_{2n-2}} \mathcal{W}_{\left(0\right) j_1 j_2}^{i_1 i_2} \cdots \mathcal{W}_{\left(0\right) j_{2n-5} j_{2n-4}}^{i_{2n-5} i_{2n-4}} \left[ \mathcal{S}_{\left(0\right)} \mathcal{W}_{\left(0\right) j_{2n-3} j_{2n-2}}^{i_{2n-3} i_{2n-2}} \right.\notag  \\
&  \left.+ 2\left(n-1\right) \left(\frac{1}{n-2} \mathcal{B}_{\left(0\right) j_{2n-3}}^{i_{2n-3}} \delta_{j_{2n-2}}^{i_{2n-2}} + \mathcal{S}_{\left(0\right)m}^{i_{2n-3}} \mathcal{W}_{\left(0\right) j_{2n-3} j_{2n-2}}^{i_{2n-2} m} + \nabla^{i_{2n-3}}_{(0)} \mathcal{C}^{i_{2n-2}}_{\left(0\right) j_{2n-3} j_{2n-2}}\right)\right] +\cdots\,.
\end{align}
This expression can be further simplified due to the identity
\begin{align}
\delta^{j_1 \cdots j_{2n-1}}_{i_1 \cdots i_{2n-1}} \mathcal{W}_{\left(0\right) j_1 j_2}^{i_1 i_2} \cdots \mathcal{W}_{\left(0\right) j_{2n-3} j_{2n-2}}^{i_{2n-3} i_{2n-2}} \mathcal{S}_{\left(0\right) j_{2n-1}}^{i_{2n-1}} &= \delta^{j_1 \cdots j_{2n-2}}_{i_1 \cdots i_{2n-2}} \mathcal{W}_{\left(0\right) j_1 j_2}^{i_1 i_2} \cdots \mathcal{W}_{\left(0\right) j_{2n-5} j_{2n-4}}^{i_{2n-5} i_{2n-4}} \times \notag \\
& \times \left[\mathcal{W}_{\left(0\right) j_{2n-3} j_{2n-2}}^{i_{2n-3} i_{2n-2}} \mathcal{S}_{\left(0\right)} + 2\left(n-1\right) \mathcal{S}_{\left(0\right)m}^{i_{2n-3}} \mathcal{W}_{\left(0\right) j_{2n-3} j_{2n-2}}^{i_{2n-2} m}\right] \,,
\end{align}
what gives rise to the Weyl variation of the  part (ii) of the form
\begin{align}
\delta_{\sigma} &I_{2n+1}^{(ii)} = -\frac{\sigma \ell n c_{2n}}{2^{n-2}}  \int\limits_{\partial
\mathcal{M}}d^{2n}x\,\sqrt{-g_{\left(0\right)}} \left[\delta^{j_1 \cdots j_{2n-1}}_{i_1 \cdots i_{2n-1}} \mathcal{W}_{\left(0\right) j_1 j_2}^{i_1 i_2} \cdots \mathcal{W}_{\left(0\right) j_{2n-3} j_{2n-2}}^{i_{2n-3} i_{2n-2}} \mathcal{S}_{\left(0\right) j_{2n-1}}^{i_{2n-1}} + \notag \right. \\
& \left. + 2\left(n-1\right) \,\delta^{j_1 \cdots j_{2n-2}}_{i_1 \cdots i_{2n-2}} \mathcal{W}_{\left(0\right) j_1 j_2}^{i_1 i_2} \cdots \mathcal{W}_{\left(0\right) j_{2n-5} j_{2n-4}}^{i_{2n-5} i_{2n-4}} \left(\frac{1}{n-2} \mathcal{B}_{\left(0\right) j_{2n-3}}^{i_{2n-3}} \delta_{j_{2n-2}}^{i_{2n-2}} + \nabla_{\left(0\right)}^{i_{2n-3}} \mathcal{C}^{i_{2n-2}}_{\left(0\right) j_{2n-3} j_{2n-2}}\right)\right] +\cdots \,.
\label{Iiiuniversal}
\end{align}
Clearly, there are polynomial and non-polynomial terms in the Weyl and Schouten tensors in this part of the Weyl variation of the total action.  While derivative contributions may be corrected by extra contributions from part (i), it is worth to emphasize that the term with $(n-2)$ Weyl and one Schouten
\begin{equation}
\delta _{\sigma } I^{(ii)}_{2n+1}= \int\limits_{\partial \mathcal{M}
}d^{2n}x\,\sqrt{-g_{(0)}}\,\sigma \,\mathcal{A}_{\mathrm{B}}^{(ii)}\,,
\end{equation}
where
\begin{equation} \label{WWS}
\mathcal{A}_{\mathrm{B}}^{(ii)} =-
\frac{ \ell n c_{2n}}{2^{n-2}}  \delta^{j_1 \cdots j_{2n-1}}_{i_1 \cdots i_{2n-1}} \mathcal{W}_{\left(0\right) j_1 j_2}^{i_1 i_2} \cdots \mathcal{W}_{\left(0\right) j_{2n-3} j_{2n-2}}^{i_{2n-3} i_{2n-2}} \mathcal{S}_{\left(0\right) j_{2n-1}}^{i_{2n-1}} +\cdots\,,
\end{equation}
is a generic contribution, which matches the results of seven and nine bulk dimensions~\cite{Jia:2021hgy}.

\subsection{Part (i): contribution  to type B anomaly}
\label{Sec:Part (i)}

Part (i) of the variation of the action can be expressed as two terms with single-parameter integrals, upon integrating by parts with respect to the covariant derivative $\nabla_{j_3}$. This is a result which derives from the fact that the part that depends on $t$ can be written as a total derivative of the type
\begin{equation*}
\frac{d }{d t}\,\left[ t\left( \frac{1}{2}\,\mathcal{R}_{j_{5}j_{6}}^{i_{5}i_{6}}-t^{2}K_{j_{5}}^{i_{5}}K_{j_{6}}^{i_{6}}+\frac{%
s^{2}}{\ell ^{2}}\,\delta _{j_{5}}^{i_{5}}\delta _{j_{6}}^{i_{6}}\right)
\cdots \left( \frac{1}{2}\,\mathcal{R}_{j_{2n-1}j_{2n}}^{i_{2n-1}i_{2n}}-t^{2}K_{j_{2n-1}}^{i_{2n-1}}K_{j_{2n}}^{i_{2n}}+
\frac{s^{2}}{\ell ^{2}}\,\delta _{j_{2n-1}}^{i_{2n-1}}\delta_{j_{2n}}^{i_{2n}}\right) \right] .
\end{equation*}
This simplification allows to write down a generic formula, where it is simpler to implement a power-counting argument, at least in one of the terms. In doing so, 
\begin{eqnarray}
\delta I_{2n+1}^{(i)} &=&2n(n-1)c_{2n}\int\limits_{\partial \mathcal{M}}d^{2n}x\,\sqrt{-h}\int\limits_{0}^{1}dt\,\delta _{i_{1}\cdots
i_{2n}}^{j_{1}\cdots j_{2n}}\,\delta _{j_{1}}^{i_{1}}\,\nabla
_{j_{3}}K_{j_{2}}^{i_{2}}\,\nabla ^{i_{3}}\left( h^{-1}\delta h\right)
_{j_{4}}^{i_{4}}\times   \notag \\
&&\qquad \qquad \times \left( \frac{1}{2}\,R_{j_{5}j_{6}}^{i_{5}i_{6}}+\frac{t^{2}}{\ell ^{2}}\,\delta _{j_{5}}^{i_{5}}\delta _{j_{6}}^{i_{6}}\right)
\cdots \left( \frac{1}{2}\,R_{j_{2n-1}j_{2n}}^{i_{2n-1}i_{2n}}+\frac{t^{2}}{%
\ell ^{2}}\,\delta _{j_{2n-1}}^{i_{2n-1}}\delta _{j_{2n}}^{i_{2n}}\right)  
\label{ipartanomaly} \\
&&-2n(n-1)c_{2n}\int\limits_{\partial \mathcal{M}}d^{2n}x\,\sqrt{-h}%
\int\limits_{0}^{1}dt\,t\,\delta _{i_{1}\cdots i_{2n}}^{j_{1}\cdots
j_{2n}}\,\delta _{j_{1}}^{i_{1}}\,\nabla _{j_{3}}K_{j_{2}}^{i_{2}}\nabla ^{i_{3}}\left(
h^{-1}\delta h\right) _{j_{4}}^{i_{4}}\,\times   \notag \\
&& \times \left( \frac{1}{2}\,\mathcal{R}%
_{j_{5}j_{6}}^{i_{5}i_{6}}-t^{2}K_{j_{5}}^{i_{5}}K_{j_{6}}^{i_{6}}+\frac{%
t^{2}}{\ell ^{2}}\,\delta _{j_{5}}^{i_{5}}\delta _{j_{6}}^{i_{6}}\right)
\cdots \left( \frac{1}{2}\,\mathcal{R}%
_{j_{2n-1}j_{2n}}^{i_{2n-1}i_{2n}}-t^{2}K_{j_{2n-1}}^{i_{2n-1}}K_{j_{2n}}^{i_{2n}}+%
\frac{t^{2}}{\ell ^{2}}\,\delta _{j_{2n-1}}^{i_{2n-1}}\delta
_{j_{2n}}^{i_{2n}}\right) \,. \notag
\end{eqnarray}
The asymptotic analysis of the first two lines has to be carried out on a case-by-case basis. Examples are provided in five and seven dimensions in Section~\ref{particular}.

In turn, the expansion in the holographic radial coordinate $z$, applied to the last two lines of the above expression leads to a finite term at the boundary. Thus, the contribution to the Weyl anomaly is given by
\begin{eqnarray}
&&-2n(n-1)c_{2n}\ell \int\limits_{\partial \mathcal{M}}d^{2n}x\,\,\sqrt{%
-g_{(0)}}\,\int\limits_{0}^{1}du\,\delta _{i_{1}\cdots i_{2n}}^{j_{1}\cdots
j_{2n}}\,\delta _{j_{1}}^{i_{1}}\delta _{j_{2}}^{i_{2}}\,\nabla _{(0)j_{3}}%
\mathcal{S}_{(0)j_{4}}^{i_{4}}\,\nabla _{(0)}^{i_{3}}\sigma \,  \notag \\
&&\times \left( \frac{1}{2}\,\mathcal{R}_{(0)j_{5}j_{6}}^{i_{5}i_{6}}-2u\,%
\delta _{j_{5}}^{i_{5}}\mathcal{S}_{(0)j_{6}}^{i_{6}}\right) \cdots \left( 
\frac{1}{2}\,\mathcal{R}_{(0)j_{2n-1}j_{2n}}^{i_{2n-1}i_{2n}}-2u\,\delta
_{j_{2n-1}}^{i_{2n-1}}\mathcal{S}_{(0)j_{2n}}^{i_{2n}}\right) \,.
\end{eqnarray}%
As the result is written in terms of covariant quantities of the boundary metric $g_{(0)ij}$, the integration by parts of  $\nabla _{(0)}^{i}\sigma $ would correspond to a type C anomaly, i.e., a trivial part in the form of a total derivative.

\section{Particular cases} 
\label{particular}

\subsection{Five dimensions}

One can show that extra contributions to the Weyl anomaly, on top of the Pfaffian of the Riemann and Weyl tensors, vanish in the five-dimensional case.

Indeed, the generic expression for the part (ii) of the Weyl variation in Eq.~\eqref{Iiipartgeneric} adopts the particular form
\begin{equation}
\delta_\sigma I_{5}^{(ii)}
= \frac{\ell^2}{128 \pi G_{N}}\int\limits_{\partial \mathcal{M}} d^4x\, 
\sqrt{-h}\,\delta^{j_1 j_2 j_3}_{i_1 i_2 i_3}
\left[\left( h^{-1}\delta_{\sigma} h\right) _{k}^{i_1}K_{j_1}^{k}+2\delta_{\sigma}
K_{j_1}^{i_1}\right] W^{i_2 i_3}_{j_2 j_3} \,,
\end{equation}
which can be readily re-written in terms of the electric part of the Weyl tensor as
\begin{equation}
\delta_\sigma I_{5}^{(ii)} 
= \frac{\ell^2}{32 \pi G_{N}}\int\limits_{\partial \mathcal{M}} d^4x\, 
\sqrt{h}\,
\left[\left( h^{-1}\delta_{\sigma} h\right) _{k}^{i}K_{j}^{k}+2\delta_{\sigma}
K_{j}^{i}\right]
W^{j z}_{i z} \,.
\end{equation}
However, the power-series expansion for the electric part of the Weyl tensor in Eq.~\eqref{electric} breaks down in four boundary dimensions. Anyhow, these components can be put back in the FG form, such that
\begin{align}
W^{iz}_{jz} = \frac{z^4}{\ell^2}\left(g^{ik}_{(2)}g_{(2)jk}-4g^{i}_{(4)j}\right) +\mathcal{O} \left(z^6\right) \,.
\end{align}
Since this part of the Weyl tensor falls-off at normalizable order, one can truncate the series in~\eqref{det_h},~\eqref{vol} and the square bracket~\eqref{dKcal} at leading order. This calculation results in the trace of the electric part of the Weyl tensor, what vanishes at all orders. In particular, at holographic order, for Einstein spacetimes, this relation implies
\begin{equation}
     g^{i}_{(4)i}= \frac{1}{4} \, g^{ik}_{(2)}g_{(2)ik} \,.
\end{equation}
In doing so, $\delta_\sigma I_{5}^{(ii)}=0$.

Since only the first two lines of Eq.~\eqref{ipartanomaly} contribute to the non-trivial part of the Weyl anomaly, for the particular case of five spacetime dimensions, it reads
\begin{align}
\delta_{\sigma} I_{5}^{(i)} 
&= \frac{\ell^2}{32 \pi G_{N}} 
\int\limits_{\partial \mathcal{M}} d^4x \,
\sqrt{h}\,
\delta_{i_1 i_2 i_3}^{j_1 j_2 j_3}
\nabla_{j_2}K_{j_1}^{i_1}
\nabla^{i_2}\left(h^{-1}\delta_{\sigma} h\right)^{i_3}_{j_3} \,.
\label{I5i}
\end{align}
By performing the asymptotic expansion of the integrand, one obtains
\begin{equation}
\delta_{i_1 i_2 i_3}^{j_1 j_2 j_3 } \, \nabla_{j_2}K_{j_1}^{i_1} =\frac{z^2}{2 \ell}\, \delta_{i_1 i_2 i_3}^{j_1 j_2 j_3 }\, \mathcal{C}^{i_1}_{\left(0\right) j_1 j_2} +\mathcal{O}\left(z^4\right)\,.
\label{deldelK}
\end{equation}
Furthermore, considering the variation of the induced metric under local Weyl rescalings of the boundary metric $g_{\left(0\right)ij}$, that expands as
\begin{equation}
\left( h^{-1}\delta_{\sigma} h\right) _{k}^{i} = 2 \sigma \delta^i_k +\mathcal{O} \left(z^2\right) \,.
\end{equation}
Eq.~\eqref{I5i} can be cast in the form below
\begin{equation}
\delta_{\sigma} I_{5}^{(i)} 
= \frac{\ell^3}{16 \pi G_{N}} 
\int\limits_{\partial \mathcal{M}} d^4x \,
\sqrt{-g_{\left(0\right)}}\,
\delta_{i_1 i_2}^{j_1 j_2}
\mathcal{C}^{i_1}_{\left(0\right) j_1 j_2}
\nabla^{i_2}_{\left(0\right)}\sigma \,,
\end{equation}
what vanishes identically due to the resulting trace of the Cotton tensor.

In summary, the only non-vanishing contribution arises from the part (iv), which involves the difference between the Pfaffian of the Riemann and Weyl tensor~\eqref{TypeA-PfB}, 
\begin{equation}
\delta_\sigma I_{5} = -\frac{\sigma \ell^3}{128 \pi G_N} \int\limits_{\partial \mathcal{M}
}d^{4}x\,\sqrt{-g_{(0)}} \left(\mathcal{W}_{\left(0\right) kl}^{ij} \mathcal{W}_{\left(0\right) ij}^{kl} - \mathcal{E}_4\right) \,,
\end{equation}
in agreement with existing literature~\cite{Henningson:1998gx}.

\subsection{Seven dimensions}

In seven dimensions, Eq.~\eqref{Iiipartgeneric} induces two contributions at the level of the variation 
\begin{equation} \label{7Dii}
\delta I_{7}^{\left(ii\right)}
= -\frac{\ell^4}{4096 \pi G_N} \int\limits_{\partial \mathcal{M}} d^6 x\, \sqrt{-h}\,
  \delta^{j_1 \ldots j_5}_{i_1 \ldots i_5} 
\left[
\left(h^{-1} \delta h\right)^{i_1}_{k} K^{k}_{j_1} + 2 \delta K^{i_1}_{j_1}
\right] W^{i_2 i_3}_{j_2 j_3}  
\left(W^{i_4 i_5}_{j_4 j_5}- \frac{4}{3\ell^2} 
\delta^{i_4 i_5}_{j_4 j_5}\right) \,.
\end{equation}
Since $\sqrt{-h}$ diverges as $\mathcal{O} \left(z^{-6}\right)$, one is aiming at contributions of the integrand that fall-off at normalizable order, i.e., $\mathcal{O} \left(z^{6}\right)$, in order to produce the finite terms.

For arbitrary variations of the fields, the second term is proportional to the electric part of the Weyl tensor
\begin{equation}
\frac{\ell^2}{64 \pi G_N}  \int\limits_{\partial
\mathcal{M}}d^{6}x\,\sqrt{-h} \,\left[ \left( h^{-1}\delta_{\sigma} h\right)
_{k}^{i}K_{j}^{k}+2\delta_{\sigma} K_{j}^{i} \right] W^{jz}_{iz} \,.
\end{equation}
As the expansion of $W^{jz}_{iz}$ starts at $\mathcal{O} \left(z^4\right)$, only the leading order of the square bracket~\eqref{dKcal} is relevant for the discussion. For a Weyl rescaling of the metric, only the trace appears up to the holographic order, what vanishes identically. In general, it can be shown that the first subleading term falls-off as $\mathcal{O} \left(z^8\right)$ and, then, it contributes to the Weyl anomaly in 9D. 

As a consequence, the only contribution to the part (ii) of the 7D Weyl anomaly comes from the first term in Eq.~\eqref{7Dii}. This piece was derived in an  arbitrary even-dimensional boundary in Eq.~\eqref{Iiiuniversal}. For this particular case, it reads
\begin{equation}\label{PartTBAnomalyv1}
\delta_{\sigma} I^{\left(ii\right)}_{7}=\frac{\sigma \ell^5}{2048 \pi G_N} \int\limits_{\partial \mathcal{M}} d^6x\,\sqrt{-g_{(0)}} \left[{\delta^{j_1 j_2 j_3 j_4j_5}_{i_1{i}_2{i}_3{i}_4{i}_5}\mathcal{S}^{i_1}_{(0)j_{1}}\mathcal{W}^{i_2 i_3}_{(0){j}_2 {j}_3}\mathcal{W}^{i_4 i_5}_{(0){j}_4 {j}_5}}  + 4 \delta^{j_1 j_2 j_3 j_4}_{i_1{i}_2{i}_3{i}_4}\mathcal{W}^{i_1 i_2}_{\left(0\right){j}_1 {j}_2} \nabla^{i_3}\mathcal{C}^{i_4}_{\left(0\right) j_3 j_4} \right]\,.
\end{equation}
The last formula can be suitably rewritten, after integrating by parts, in terms of the Cotton tensor as
\begin{equation}\label{PartTBAnomaly}
\delta_{\sigma} I^{\left(ii\right)}_{7}=\frac{\sigma \ell^5}{2048 \pi G_N}\int\limits_{\partial \mathcal{M}} d^6x\,\sqrt{-g_{\left(0\right)}} \left({\delta^{j_1 j_2 j_3 j_4j_5}_{i_1{i}_2{i}_3{i}_4{i}_5}\mathcal{S}^{i_1}_{\left(0\right)j_{1}}\mathcal{W}^{i_2 i_3}_{\left(0\right){j}_2 {j}_3}\mathcal{W}^{i_4 i_5}_{\left(0\right){j}_4 {j}_5}} + 48 \mathcal{C}_{\left(0\right)i}^{jk}\mathcal{C}^{i}_{\left(0\right)jk} \right)\,.
\end{equation}
The total derivative coming from the integration by parts contributes to the type C anomaly, such that it may be consistently dropped out.

The part (i), in Eq.~\eqref{ipartanomaly}, is conveniently cast in a form where a power-counting analysis can be readily taken. In particular, only the first two lines contribute to the Weyl anomaly. When  Weyl rescalings are considered, this part of the variation reads
\begin{equation}
\delta_{\sigma} I_{7}^{\left(i\right)} = -\frac{\ell^4}{512 \pi G_N} \int\limits_{\partial
\mathcal{M}} d^6 x \,\sqrt{-h} \int\limits_{0}^{1}dt \delta^{j_1 \cdots j_5}_{i_1 \cdots i_5} \nabla_{j_2}K^{i_1}_{j_1} \nabla^{i_2}\left(h^{-1} \delta_\sigma h\right)^{i_3}_{j_3}\left(
R^{{i}_4 {i}_5}_{{j}_4 {j}_5}
+ \frac{2t^2}{\ell^2} \delta^{{i}_4}_{{j}_4} \delta^{{i}_5}_{{j}_5}\right)  \,.
\end{equation}
Moreover, for the variation of the induced metric under Weyl rescaling of $g_{\left(0\right)ij}$ we get
\begin{equation}
\left( h^{-1}\delta_{\sigma} h\right) _{k}^{i} = 2 \sigma \left(\delta^i_k +z^2 \mathcal{S}^i_{\left(0\right)k}+\mathcal{O} \left(z^4\right)\right) \,.
\end{equation}
Notice that the leading order of this term contributes solely to the Type C anomaly, hence, it can be omitted. Thus, only the subleading order is relevant for this computation. Indeed, taking into account Eq.~\eqref{deldelK}, the derivative contributions of the integrand are of normalizable order--$\mathcal{O} \left(z^6\right)$. Namely, only the leading order contributions of the rest of the term are relevant for the 7D Weyl anomaly. Overall, the finite contribution can be cast in the form
\begin{equation}
\delta_{\sigma} I_{7}^{\left(i\right)}= -\frac{\sigma \ell^5}{128 \pi G_N} \int_{\partial \mathcal{M}} d^6x\,\sqrt{-g_{\left(0\right)}} \, \delta^{j_1 j_2 j_3 }_{i_1 i_2 i_3} \mathcal{C}^{i_1}_{\left(0\right) j_1 j_2} \mathcal{C}^{i_2 i_3}_{\left(0\right) j_3}  \,,
\end{equation}
or equivalently
\begin{equation}\label{iv-7}
\delta_{\sigma} I^{\left(i\right)}_{7}
=-\frac{\sigma \ell^5}{64 \pi G_N}\int\limits_{\partial
\mathcal{M}} d^6 x \,\sqrt{-g_{(0)}} \, \mathcal{C}^{i }_{\left(0\right) jm} \mathcal{C}^{jm} _{\left(0\right)i} \,.
\end{equation}
In conclusion, taking into account Eqs.~\eqref{TypeA-PfB},~\eqref{PartTBAnomaly} and~\eqref{iv-7}, we obtain
\begin{align}
\delta_\sigma I_7 = \frac{\sigma \ell^5}{24576 \pi G_N} \int\limits_{\partial
\mathcal{M}} d^6 x \,\sqrt{-g_{(0)}} &\left[ \delta^{i_1 \ldots i_6}_{j_1 \ldots j_6} \left(\mathcal{W}^{j_1 j_2}_{ \left(0\right) i_1 i_2} \mathcal{W}^{j_3 j_4}_{ \left(0\right) i_3 i_4} \mathcal{W}^{j_5 j_6}_{ \left(0\right) i_5 i_6} - \mathcal{R}^{j_1 j_2}_{ \left(0\right) i_1 i_2} \mathcal{R}^{j_3 j_4}_{ \left(0\right) i_3 i_4} \mathcal{R}^{j_5 j_6}_{ \left(0\right) i_5 i_6}\right) \nonumber \right. \\
& \left. + 12 \delta^{i_1 \ldots i_5}_{j_1 \ldots j_5} \mathcal{W}^{j_1 j_2}_{ \left(0\right) i_1 i_2} \mathcal{W}^{j_3 j_4}_{ \left(0\right) i_3 i_4} \mathcal{S}^{j_5}_{\left(0\right)i_5} + 192 \mathcal{C}_{\left(0\right)j k}^{i} \mathcal{C}_{\left(0\right)i}^{jk}\right] \,,
\end{align}
that matches the result in the literature for the 7D anomaly~\eqref{7Danomalykarydas}.

\section{Conclusions}

In this work,  a prescription on how to obtain holographic information on conformal anomalies, dual to Einstein-AdS gravity, is provided in all odd dimensions. This is carried out within the framework of a renormalization scheme for AdS gravity, which considers the addition of extrinsic counterterms. While some divergent terms in the variation of the action may remain, they are given in terms of tensors which account for conformal properties of the metric $g_{(0)ij}$ and, therefore, they are identically zero for conformally flat boundaries. In the general case, extra counterterms should be added to the gravitational action, which do not modify universal holographic quantities of the theory as, e.g.,  conformal anomalies. 

Therefore, despite the mismatch in the boundary counterterms with respect to Holographic Renormalization, Kounterterms method is able to determine a substantial part of the holographic anomalies in any $2n+1$ dimension:

\begin{itemize}
    
\item the central charge $a$  of type A anomaly,

\item  the central charge $c$ of the Pfaffian of the Weyl tensor, which turns out to be $c=a$ for boundary field theories dual to Einstein gravity,

\item  the coefficient of the totally-antisymmetric product between $(n-1)$ Weyls and a single Schouten tensor in Eq.~\eqref{WWS}.
\end{itemize}

 Other contributions to the type B trace anomaly can be worked out in particular dimensions from the generic formulae. 
 
 The generality of the above results is a consequence of a variation of the action which is polynomial in the extrinsic and intrinsic curvatures. Indeed, just the lowest-order modes in the FG expansion are required in order to express the variation in terms of $2n$ derivatives of the conformal data $g_{(0)ij}$. This reasoning also justifies the fact Log terms play no role in the above generic computation. This is in stark contrast with standard techniques in Einstein-AdS gravity, where the obtention of anomalies requires the asymptotic solution  of the canonical momentum --which is linear in the extrinsic curvature-- until the holographic order. Within that framework, the addition the counterterms only cancels the corresponding divergences at the conformal boundary, such that they do not alter the holographic correlators of the gravity theory. 
 
 For pure Einstein-AdS gravity, the sign of the vacuum energy alternates with the value of $n$ in $2n$ boundary dimensions. In particular, the vacuum energy for the ground state of AdS$_{3}$ gravity is that where the BTZ
mass is $M=-1$~\cite{Balasubramanian:1999re}. As the gauge/gravity duality implies a correspondence between the energy of global AdS with the Casimir energy of the dual CFT$_{2n}$ on $\mathbb{R} \times \mathbb{S}^{2n-1}$, the alternating sign is reflected in the central charges $a$ and $c$~\cite{Assel:2015nca}. 

Kounterterms at the boundary of odd-dimensional AAdS spaces, as a mathematical structure, are closely related to extensions of Chern-Simons densities, known as Transgression Forms~\cite{Mora:2006ka}. They are intended to accommodate an additional gauge connection in the algebra, such that the new object is truly gauge invariant. It would be interesting to explore the connection of these structures to the ones appearing in other  frameworks in the existing literature, where the conformal anomaly is derived in various dimensions~\cite{Imbimbo:2023sph,apruzzi2025symmetrytftscontinuousspacetime,aminov2026typeaconformalanomalieseuler}.

\section*{Acknowledgements}

We thank I.J.~Araya, N.~Boulanger, C.~Corral, I.~García-Etxebarria and P.~Sundell for interesting discussions. J.B.C.~thanks the Departamento de Ciencias at Universidad Adolfo Ibáñez for their hospitality during this work. The work of J.B.C.~is supported by Becas de Postgrado UNAP. This work was partially funded by Agencia Nacional de Investigación y Desarrollo (ANID) through the FONDECYT Grants No.~11240059, 1240043, 1230492, 1231779, 1240955 and 1261016.

\section*{Data Availiabilty Statement}
This work is purely theoretical and does not involve the generation or analysis of empirical datasets. All results are derived analytically and are fully contained within the article. No additional data are required to support the findings of this study.

\appendix

\section{Conventions}
\label{GC}

For the radial evolution of the spacetime geometry, the line element is expressed as
\begin{equation}
ds^2 = N^2(z) d{z^2} + h_{ij} \left(z, x\right) d{x^i} d{x^j} \,,
\end{equation}
where $N(z)$ is the lapse function. The normal to the constant $z$ surfaces takes the form
\begin{equation}
    n_{\mu} = -N \delta^z_{\mu} \,.
\end{equation}
In this coordinate frame, the extrinsic curvature adopts a simpler form with respect to the standard ADM foliation, that is,
\begin{equation}
K_{ij} = - \frac{1}{2N} \, \partial_z h_{ij} \,.
\end{equation}
The independent, non-vanishing components of the Christoffel symbols are written as
\begin{equation}
\Gamma^z_{zz} = \frac{N'}{N} \,, \quad \Gamma^z_{ij} = \frac{1}{N} \, K_{ij} \,, \quad \Gamma^{i}_{zj} = -N K_j^i \,, \quad \Gamma^i_{jk}\left(g\right) = \Gamma^i_{jk} \left(h\right) \,.
\end{equation}
The above expressions lead to the Gauss-Codazzi relation for a radial foliation of the spacetime
\begin{align}
R^{ij}_{km} &=\mathcal{R}^{ij}_{km} \left(h\right)- \left(K^i_k K^j_m -K^i_m K^j_k\right) \,, \\
R^{iz}_{mz}&=-\frac{1}{N} \, \partial_z K^i_m - K^i_s K^s_m \,, \\
R^{iz}_{jm}&=-\frac{1}{N} \, \left(\mathcal{D}_j K^i_m - \mathcal{D}_m K^i_j\right) \,, \\
R&=\mathcal{R}\left(h\right) - \left(K^2 + K^i_j K^j_i\right) -\frac{2}{N}\, \partial_z K \,.
\end{align}



The totally antisymmetric Kronecker delta of rank $p\leq d$ in $d$ dimensions is defined as the determinant of the $p\times p$ matrix whose entries are $\delta _{j_{l}}^{i_{m}}$, with $m,l=1,\ldots ,p$. A contraction
of $k\leq p$ pairs of indices produces a delta of rank $p-k$, 
\begin{equation}
\delta _{i_{1}\cdots i_{k}\cdots i_{p}}^{j_{1}\cdots j_{k}\cdots j_{p}}\,\delta _{j_{1}}^{i_{1}}\cdots \delta _{j_{k}}^{i_{k}}=\frac{\left(
d-p+k\right) !}{\left( d-p\right) !}\,\delta _{i_{k+1}\cdots i_{p}}^{j_{k+1}\cdots j_{p}}\,.
\end{equation}

\section{Variation of Einstein-AdS action plus Kounterterms} 
\label{AP:Variation}

The variation of Einstein-AdS action plus Kounterterms discussed in Sec.~\ref{Sec:Variation}  was derived in Refs.~\cite{Olea:2006vd,Miskovic:2008ck}. The purpose of this Appendix is to isolate the part of the variation of the boundary contribution $B_{2n}$ which depends on variations of the induced metric. To this aim, the variation  $\delta (\sqrt{-h})=\frac{1}{2}\,\sqrt{-h}\,(h^{-1}\delta h)_{k}^{k}$
gets combined with other of the form $\delta h^{i_{4}k}\mathcal{R}_{\ kj_{3}j_{4}}^{i_{3}}$ to produce
\begin{eqnarray}
&&-nc_{2n}\int\limits_{\partial \mathcal{M}}d^{2n}x\,\sqrt{-h}%
\,\int\limits_{0}^{1}dt\int\limits_{0}^{t}ds\,\delta _{i_{1}\cdots
i_{2n}}^{j_{1}\cdots j_{2n}}\,K_{j_{1}}^{i_{1}}\delta _{j_{2}}^{i_{2}}\times \notag
\\
&&\times \left[ (h^{-1}\delta h)_{k}^{k}\,\left( \frac{1}{2}\,\mathcal{R}_{j_{2}j_{4}}^{i_{3}i_{4}}-t^{2}K_{j_{3}}^{i_{3}}K_{j_{4}}^{i_{4}}+\frac{s^{2}}{\ell ^{2}}\,\delta _{j_{3}}^{i_{3}}\delta _{j_{4}}^{i_{4}}\right)
-(n-1)\,(h^{-1}\delta h)_{k}^{i_{3}}\mathcal{R}_{j_{3}j_{4}}^{ki_{4}}\right]
\times \\
&&\times \left( \frac{1}{2}\,\mathcal{R}_{j_{5}j_{6}}^{i_{5}i_{6}}-t^{2}K_{j_{5}}^{i_{5}} K_{j_{6}}^{i_{6}}+\frac{s^{2}}{\ell ^{2}}\,\delta _{j_{5}}^{i_{5}}\delta _{j_{6}}^{i_{6}}\right)
\cdots \left( \frac{1}{2}\,\mathcal{R}_{j_{2n-1}j_{2n}}^{i_{2n-1}i_{2n}}-t^{2}K_{j_{2n-1}}^{i_{2n-1}}K_{j_{2n}}^{i_{2n}}+
\frac{s^{2}}{\ell ^{2}}\,\delta _{j_{2n-1}}^{i_{2n-1}}\delta
_{j_{2n}}^{i_{2n}}\right) . \notag
\end{eqnarray}
The trace $(h^{-1}\delta h)_{k}^{k}$ can be traded off by a sum of index insertions using the permutational identity 
\begin{equation*}
0=\delta _{i_{1}\cdots i_{2n}i_{2n+1}}^{j_{1}\cdots j_{2n}j_{2n+1}}=\delta
_{i_{1}}^{j_{1}}\,\delta _{i_{2}\cdots i_{2n+1}}^{j_{2}\cdots
j_{2n+1}}-\delta _{i_{1}}^{j_{2}}\,\delta _{i_{2}\cdots
i_{2n+1}}^{j_{1}j_{3}\cdots j_{2n+1}}+\cdots +\,\delta
_{i_{1}}^{j_{2n+1}}\,\delta _{i_{2}\cdots i_{2n+1}}^{j_{2}\cdots
j_{2n}}=\sum_{p=1}^{2n+1}(-1)^{p-1}\,\delta _{i_{1}}^{j_{p}}\,\delta
_{i_{2}\cdots i_{2n+1}}^{j_{2}\cdots j_{p-1}j_{p+1}\cdots j_{2n+1}}
\end{equation*}
which follows from the vanishing of the totally anti-symmetric Kronecker delta of rank $2n+1$ in $2n$ dimensions.

Contracting this expression with $(h^{-1}\delta h)_{j_{2n+1}}^{i_{2n+1}}$ yields
\begin{equation}
\delta _{i_{1}\cdots i_{2n}}^{j_{1}\cdots j_{2n}}\,(h^{-1}\delta
h)_{k}^{k}=\sum_{p=1}^{2n}\delta _{i_{1}\cdots i_{p-1}ki_{p+1}\cdots
i_{2n}}^{j_{1}\cdots j_{2n}}(h^{-1}\delta h)_{i_{p}}^{k}\,.
\end{equation}
Upon further contraction with all the pairs of indices in $K\,\delta \left( \frac{1}{2}\,\mathcal{R}-t^{2}K^{2}+\frac{t^{2}}{\ell ^{2}}\,\delta ^{2}\right)
^{n-2}$, the first two terms in the sum reorganize into the combination  proportional to $(h^{-1}\delta h)_{k}^{i_{1}}\left( K_{j_{1}}^{k}\delta_{j_{2}}^{i_{2}}-\delta _{j_{2}}^{k}K_{j_{1}}^{i_{2}}\right) $, while the remaining $2n-2$ terms become identical after dummy-index relabeling. Then, the
result is
\begin{eqnarray}
&&-nc_{2n}\int\limits_{\partial \mathcal{M}}d^{2n}x\,\sqrt{-h}%
\,\int\limits_{0}^{1}dt\int\limits_{0}^{t}ds\,\delta _{i_{1}\cdots
i_{2n}}^{j_{1}\cdots j_{2n}}\,(h^{-1}\delta h)_{k}^{i_{1}}\left[
\rule[2pt]{0pt}{14pt}\left( K_{j_{1}}^{k}\delta _{j_{2}}^{i_{2}}-\delta_{j_{2}}^{k}K_{j_{1}}^{i_{2}}\right) \times \right.   \notag \\
&&\times \left( \frac{1}{2}\,\mathcal{R}_{j_{2}j_{4}}^{i_{3}i_{4}}-t^{2}K_{j_{3}}^{i_{3}}K_{j_{4}}^{i_{4}}+\frac{%
s^{2}}{\ell ^{2}}\,\delta _{j_{3}}^{i_{3}}\delta _{j_{4}}^{i_{4}}\right)
+\left. \rule[2pt]{0pt}{12pt}(n-1)\left(
-t^{2}K_{[j_{1}}^{k}K_{j_{2}]}^{i_{2}}+\frac{s^{2}}{\ell ^{2}}\,\delta
_{\lbrack j_{1}}^{k}\delta _{j_{2}]}^{i_{2}}\right) K_{j_{3}}^{i_{3}}\delta
_{j_{4}}^{i_{4}}\right] \times   \notag \\
&&\times \left( \frac{1}{2}\,\mathcal{R}_{j_{5}j_{6}}^{i_{5}i_{6}}-t^{2}K_{j_{5}}^{i_{5}}K_{j_{6}}^{i_{6}} +\frac{s^{2}}{\ell ^{2}}\,\delta _{j_{5}}^{i_{5}}\delta _{j_{6}}^{i_{6}}\right)
\cdots \left( \frac{1}{2}\,\mathcal{R}%
_{j_{2n-1}j_{2n}}^{i_{2n-1}i_{2n}}-t^{2}K_{j_{2n-1}}^{i_{2n-1}}K_{j_{2n}}^{i_{2n}}+%
\frac{s^{2}}{\ell ^{2}}\,\delta _{j_{2n-1}}^{i_{2n-1}}\delta
_{j_{2n}}^{i_{2n}}\right) \,.  \notag
\end{eqnarray}
From this point, the remaining steps are direct: the double integrals in $t$ and $s$ are reduced to a single parametric integral by rewriting the integrand as a difference of total derivatives,
\begin{eqnarray*}
&&-nc_{2n}\int\limits_{\partial \mathcal{M}}d^{2n}x\,\sqrt{-h}%
\,\int\limits_{0}^{1}dt\int\limits_{0}^{t}ds\,\delta _{i_{1}\cdots i_{2n}}^{j_{1}\cdots j_{2n}}\,(h^{-1}\delta h)_{k}^{i_{1}}\times  \\
&&\times \left\{ \frac{\partial }{\partial t}\left[ t\,K_{j_{1}}^{k}\delta
_{j_{2}}^{i_{2}}\,\left( \frac{1}{2}\,\mathcal{R}_{j_{2}j_{4}}^{i_{3}i_{4}}-t^{2}K_{j_{3}}^{i_{3}}K_{j_{4}}^{i_{4}}+\frac{s^{2}}{\ell ^{2}}\,\delta _{j_{3}}^{i_{3}}\delta _{j_{4}}^{i_{4}}\right)
\cdots \left( \frac{1}{2}\,\mathcal{R}%
_{j_{2n-1}j_{2n}}^{i_{2n-1}i_{2n}}-t^{2}K_{j_{2n-1}}^{i_{2n-1}}K_{j_{2n}}^{i_{2n}}+
\frac{s^{2}}{\ell ^{2}}\,\delta _{j_{2n-1}}^{i_{2n-1}}\delta
_{j_{2n}}^{i_{2n}}\right) \right] \right.  \\
&&-\left. \frac{\partial }{\partial s}\left[ s\,\delta
_{j_{2}}^{k}K_{j_{1}}^{i_{2}}\,\left( \frac{1}{2}\,\mathcal{R}%
_{j_{2}j_{4}}^{i_{3}i_{4}}-t^{2}K_{j_{3}}^{i_{3}}K_{j_{4}}^{i_{4}}+\frac{%
s^{2}}{\ell ^{2}}\,\delta _{j_{3}}^{i_{3}}\delta _{j_{4}}^{i_{4}}\right)
\cdots \left( \frac{1}{2}\,\mathcal{R}_{j_{2n-1}j_{2n}}^{i_{2n-1}i_{2n}}-t^{2}K_{j_{2n-1}}^{i_{2n-1}}K_{j_{2n}}^{i_{2n}}+
\frac{s^{2}}{\ell ^{2}}\,\delta _{j_{2n-1}}^{i_{2n-1}}\delta
_{j_{2n}}^{i_{2n}}\right) \right] \right\} .
\end{eqnarray*}
Performing the parametric integrations in $t$  and $s$ reconstructs the full contribution $\delta I_{2n+1}^{(iii)}$, together with the term proportional to $(h^{-1}\delta h)_{k}^{i_{1}}K_{j_{1}}^{k}$ inside $\delta I_{2n+1}^{(ii)}$ that comes from $B_{2n}$.


\bibliographystyle{ieeetr}
\bibliography{KTbiblio}

\end{document}